\documentclass[aps,prl,twocolumn,superscriptaddress,nofootinbib]{revtex4-2}

\usepackage{amsmath,amssymb}
\usepackage{graphicx}
\usepackage{bm}
\usepackage{mathrsfs}
\usepackage[none]{hyphenat}
\usepackage[colorlinks=true,linkcolor=blue,citecolor=blue,urlcolor=magenta]{hyperref}

\newcommand{\dd}{\mathrm{d}}
\newcommand{\ii}{\mathrm{i}}
\renewcommand{\Re}{\operatorname{Re}}
\renewcommand{\Im}{\operatorname{Im}}
\newcommand{\Res}{\operatorname*{Res}}
\newcommand{\FP}{\operatorname*{FP}}
\newcommand{\G}{\mathcal{G}}          % kernel
\newcommand{\cS}{\mathcal{S}}         % total exponent
\newcommand{\cM}{\mathcal{M}}         % squeezed coefficients
\newcommand{\cT}{\mathcal{T}}
\newcommand{\cN}{\mathcal{N}}
\newcommand{\Lint}{\mathcal{L}_{\rm int}}
\newcommand{\F}{\mathfrak{F}}
\newcommand{\San}{S_{\rm an}}
\newcommand{\half}{\tfrac{1}{2}}
\newcommand{\thalf}{\tfrac{3}{2}}
\newcommand{\pid}{\dot{\pi}}
\newcommand{\bk}{\mathbf{k}}
\newcommand{\bl}{\bm{\ell}}
\newcommand{\EM}[1]{\hyperref[app:#1]{End Matter~#1}}

\begin{document}

\title{Massive de Sitter Correlators as Finite Mellin--Barnes Integrals}

\author{Amara McCune}
\email{amaramccune@gmail.com}
\affiliation{Anthropic STEM Fellows Program, San Francisco, CA}

\begin{abstract}
Using the Mellin--Barnes representation of de Sitter propagators, we reduce tree-level heavy-particle exchange to finite integrals that converge at physical kinematics on mass-independent contours, and one-loop diagrams to the same integrand times a flat-space momentum factor. We compute the complete one-loop triangle at general masses, with its collider signal in closed form to all orders in the squeezed-limit expansion, and give the triple-exchange bispectrum at arbitrary heavy masses as a double series that converges at every physical shape. The loop reduction also applies at strong mixing.
\end{abstract}

\maketitle

% ===========================================================================
\textit{Introduction.}---Massive fields present during inflation imprint oscillations on primordial correlation functions, turning the early universe into a particle detector at energies near the Hubble scale~\cite{Chen2010,ArkaniHamed2015}. The oscillation frequency measures the mass, the angular dependence the spin~\cite{ArkaniHamed2015,Lee2016}, and the amplitude is Boltzmann suppressed at the Gibbons--Hawking temperature $T=H/2\pi$~\cite{Gibbons1977}. Proposed targets range from the Standard Model and its extensions~\cite{Chen2017SMbg,Chen2017SMmass,Lu2020,Hook2020,Kumar2018,Kumar2019,Craig2024,KumarNee2026,BaumannGreen2012,Chen2018nu,CuiXianyu2022} to mechanisms that lift the Boltzmann suppression or enhance the signal~\cite{WangXianyu2020,Bodas2021,Chen2022classical,KumarSundrum2020}. These signals have been searched for in the \textit{Planck} bispectrum~\cite{Sohn2024,Kumar2026planck,Kumar2026scalars,Cassem2026} and in the BOSS galaxy bispectrum~\cite{Cabass2025}, and 21-cm surveys will extend the reach~\cite{Meerburg2017}.

Which diagrams carry the signal depends on how the heavy field couples to the inflaton. A heavy singlet scalar that mixes linearly with it, as in quasi-single-field inflation, can signal most strongly through the triple exchange, and through the double exchange when its single-exchange vertex is suppressed~\cite{Chen2010,Kumar2018,Kumar2026planck,Kumar2026scalars}. Under weak mixing these are also the signals that perturbativity leaves within reach of present data at low mass~\cite{Philcox2026lim,Philcox2026dissect}. When the mixing is strong, $\rho\gtrsim H$, the quadratic theory must be solved exactly~\cite{An2018nonpert,Iyer2018strong,Huenupi2026linear,Wang2026strong,Huenupi2026,Pinol2026,Pinol2026strong}, and in the strongly mixed limit tree-level signals pay only half the usual Boltzmann suppression~\cite{Belrhali2026strong,Pinol2026strong}. Fields charged under an unbroken symmetry (including Standard Model fields while electroweak symmetry is unbroken~\cite{Chen2017SMmass}) cannot mix and first signal at one loop~\cite{Lu2020,WangXianyuZhong2022,Herderschee2026}.

Tree-level single exchange is solved in closed form~\cite{ArkaniHamed2020,QinXianyu2023closed,Belrhali2026laplace}, with singularities at mass-independent locations~\cite{Baumann2026diffeq,Gasparotto2025} whose discontinuities give the collider signal~\cite{Tong2022,QinXianyu2023loop}. Beyond single exchange, results exist diagram by diagram, except for a single heavy field, where every tree-level bispectrum reduces at any mixing to one integral over dressed legs~\cite{Pinol2026strong}. The double exchange has been computed~\cite{Aoki2024,Belrhali2026klf,Kumar2026scalars}, and its squeezed signal is known at unequal masses to leading order in the Boltzmann factor~\cite{Aoki2024}. The equal-mass triple exchange is known in the squeezed limit~\cite{Kumar2018,Chen2017SK} and in full shape~\cite{Kumar2026planck,Kumar2026scalars,Pinol2026strong}, with its squeezed coefficient in closed form at weak mixing~\cite{Pinol2026strong}. It is also known as a hypergeometric series that does not converge at physical kinematics~\cite{XianyuZang2026} and, at any mixing, as series that converge there~\cite{ChenXianyu2026}. For general trees, multivariate series exist that converge only in restricted kinematic regions~\cite{XianyuZang2024,LiuXianyu2025,Grafe2026gluing}. At one loop the bubble is under control~\cite{WangXianyuZhong2022,XianyuZhang2023,Liu2025dispersive,Qin2025loop,Zhang2026dimreg,Aoki2026fermion,Bodas2026charged,GrafeSachs2026,Chen2026IBP,Liu2026spectral}, while the triangle at general mass is known analytically only in its soft limits~\cite{QinXianyu2023loop,QinXianyu2024nonan,You2026,QinXianyu2026triangle} and numerically for a single field in the loop~\cite{Herderschee2026}. At strong mixing, the signals computed so far are all at tree level~\cite{Wang2026strong,Huenupi2026,Pinol2026,Pinol2026strong,Belrhali2026strong,Huenupi2026sound,ChenXianyu2026}.

In this \textit{Letter} we show that one reduction evaluates trees and one-loop diagrams of heavy exchange at physical kinematics, on contours that do not depend on the heavy masses. Here we apply it to the complete one-loop triangle at general masses and give its collider signal in closed form to all orders in the squeezed-limit expansion. The first correction is about half the leading term at the equilateral point, so the leading term describes the loop signal only deep in the squeezed limit. At unequal masses a second family, at the difference frequency, is comparable to the pair signal. As a test on trees, we also evaluate the triple exchange, the shape of the \textit{Planck} search of Ref.~\cite{Kumar2026planck}, at arbitrary heavy masses and leading order in the mixing. At unequal masses we give a double series that converges at every physical shape, obtained independently of that of Ref.~\cite{Aoki2026triple}.

The reduction builds on the Mellin--Barnes representation of the propagator due to Sleight and Taronna~\cite{Sleight2020,SleightTaronna2020,SleightTaronna2021a,SleightTaronna2021b,Abhishek2026}, Eq.~(\ref{eq:kernel}), which stores the time ordering of each line in one pair of Gamma functions, and on the vertex rule of the partial Mellin--Barnes method~\cite{XianyuZang2024,QinXianyu2022phase,QinXianyu2023helical}, which does the time integral at each vertex in closed form. Together they write every tree as one integral of Gamma functions over the Mellin variables, Eq.~(\ref{eq:tree}), and at one loop the loop momentum enters only through a flat-space integral with complex propagator powers. Methods that integrate out each mixing vertex, through dressed legs or mixed propagators~\cite{Chen2017SK,Pinol2026strong,ChenXianyu2026}, reduce tree-level bispectra to a single time integral. In a loop, however, the lines meet at vertices that share the loop momentum and cannot be dressed into legs. New here is the contour analysis that makes this integral usable at physical kinematics. Contours that do not pinch on the principal series are known for contact and single-exchange diagrams~\cite{Sleight2020,SleightTaronna2020}. We show that straight contours with absolute convergence exist for every tree with positive vertex powers (with a separate rule for vertices without external legs~\cite{SM}) and that they do not depend on the heavy masses. We also show how the poles of the loop's momentum factor are crossed. The integral can then be evaluated by quadrature directly at physical kinematics, and closing its contours in a fixed order gives the series and closed forms~\cite{SM}. Because the loop momentum sees each line only through a power of its energy, the same reduction applies to lines dressed by strong mixing. With the exact modes of Ref.~\cite{Belrhali2026strong} it gives the triangle's pair signal there, to our knowledge the first loop computed in that regime.

\begin{figure}[t]
\centering
\includegraphics[width=\columnwidth]{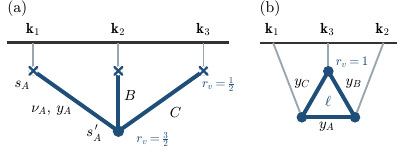}
\caption{(a) The representation read off a tree diagram, here the triple exchange. Inflaton legs are gray. Massive lines (dark blue) carry the order $\nu_l$, the energy $y_l$, and one Mellin variable at each end. Crosses are $\pid\sigma_l$ mixings ($r_v=\half$), and the filled dot is the cubic vertex $\sigma_A\sigma_B\sigma_C$ ($r_v=\thalf$). (b) The triangle, labeled for conformally coupled legs ($r_v=1$). A $\pid$ leg raises each $r_v$ to 2. The $\bl$ integral gives the momentum factor $C_3$ of Eq.~(\ref{eq:C3}).}
\label{fig:diagrams}
\end{figure}

% ===========================================================================
\textit{The representation.}---We set the Hubble rate $H$ to one. We compute the time integrals from which equal-time correlators are assembled, which have the form of connected late-time wavefunction coefficients~\cite{ArkaniHamed2017,Goodhew2021} (\EM{A}). Consider a tree of $n$ vertices $v$, in which each internal line $l$ joins vertices $v_l$, $v_l'$ and carries a scalar field of arbitrary mass $m_l$ [Fig.~\ref{fig:diagrams}]. External legs end on the late-time boundary at conformal time $\eta_0\to0$. The total energy of the legs at vertex $v$, the sum of their momentum magnitudes, is $x_v$, and the energy of line $l$ is $y_l$. On the rotated time contour $\eta=\ii t$, $t>0$, where the Bunch--Davies integrand decays as $e^{-x_v t_v}$, the coefficient with the couplings and line normalizations stripped off (\EM{A}) is
\begin{align}
\psi_n &= \int_0^\infty \prod_{v=1}^{n} \dd t_v\, t_v^{\,r_v-1}\, e^{-x_v t_v} \prod_l \G_{\nu_l}(y_l;t_{v_l},t_{v_l'})\,, \label{eq:psin}\\
\G_\nu(y;t,t') &= K_\nu(y\,t_>)\,I_\nu(y\,t_<)\,, \qquad t_>=\max(t,t')\,. \nonumber
\end{align}
Here $t_<=\min(t,t')$. The kernel $\G_\nu$ is the Euclidean anti--de Sitter propagator under the rotation $\eta=\ii t$~\cite{SleightTaronna2020,SleightTaronna2021a,DiPietro2022}. Its order carries the mass through $\nu^2=9/4-m^2$, and a heavy field has $\nu=\ii\mu$ with $\mu$ real, the principal series of de Sitter representations. The power $t_v^{\,r_v-1}$ collects the measure $a^4=t_v^{-4}$, a factor $t_v^{3/2}$ from each internal line at $v$, and a factor $t_v$ from each conformally coupled leg, so $r_v=\thalf\deg(v)+n_v-3$ for $\deg(v)$ lines and $n_v$ legs at $v$, independent of every mass. A massless leg coupled through $\pid$, as in the effective theory of inflation, contributes $t^2e^{-kt}$ in place of $t\,e^{-kt}$ and raises $r_v$ by one.

The kernel does not factorize into functions of $t$ and $t'$, and its ordering splits the integration domain into $2^L$ sectors for $L$ internal lines. One identity removes both obstructions by trading the ordered Bessel pair for Mellin variables $(s,s')$,
\begin{align}
K_\nu(y\,t_>)\,I_\nu(y\,t_<) &= \frac12\iint\!\frac{\dd s\,\dd s'}{(2\pi\ii)^2}\,
\frac{\Gamma(-s)\Gamma(-s')\Gamma(\delta)\Gamma(1-\delta)}{\Gamma(\nu+s+1)\Gamma(\nu+s'+1)} \nonumber\\
&\quad\times\Bigl(\frac{y}{2}\Bigr)^{-2\delta} t^{\nu+2s}\,t'^{\,\nu+2s'}\,, \label{eq:kernel}
\end{align}
with $\delta=-\nu-s-s'$. It is the propagator representation of Refs.~\cite{SleightTaronna2020,SleightTaronna2021a,SleightTaronna2021b,Abhishek2026} after a change of variables (\EM{A}). The two times now enter as pure powers, and the ordering is stored in $\Gamma(\delta)\Gamma(1-\delta)$, whose residues rebuild the kernel for either sign of $t-t'$. The integral converges absolutely for $\Re s,\Re s'<0$, $0<\Re\delta<\half$~\cite{SM}, and since $\Re\nu=0$ on the principal series one choice of contours serves both orderings and every mass $m>\thalf$.

With Eq.~(\ref{eq:kernel}) inserted, each $t_v$ appears only in the factor of its own vertex. Line $l$ carries $s_l$ at $v_l$ and $s_l'$ at $v_l'$, and $\Sigma_v$ is the sum of $\nu_l+2s_l$ or $\nu_l+2s_l'$ over the line ends at $v$. The vertex integral is then the single Gamma function $\Gamma(r_v+\Sigma_v)\,x_v^{-(r_v+\Sigma_v)}$~\cite{XianyuZang2024,QinXianyu2022phase,QinXianyu2023helical}, and an arbitrary tree becomes
\begin{align}
\psi_n &= \prod_l \frac12\iint\!\frac{\dd s_l\dd s_l'}{(2\pi\ii)^2}\,
\frac{\Gamma(-s_l)\Gamma(-s_l')\Gamma(\delta_l)\Gamma(1-\delta_l)}{\Gamma(\nu_l{+}s_l{+}1)\Gamma(\nu_l{+}s_l'{+}1)}\Bigl(\frac{y_l}{2}\Bigr)^{-2\delta_l} \nonumber\\
&\quad\times\prod_v \Gamma(r_v+\Sigma_v)\,x_v^{-(r_v+\Sigma_v)}\,, \label{eq:tree}
\end{align}
with $\delta_l=-\nu_l-s_l-s_l'$, a single $2L$-fold integral specified by the vertex powers $r_v$ and the line orders $\nu_l$ alone. The vertex factors decay exponentially along the contours and widen the window to $\Re s_l,\Re s_l'<0$, $0<\Re\delta_l<1$, with $\Re(r_v+\Sigma_v)>0$. On the principal series none of these conditions involves the mass, and straight contours inside the window exist whenever every $r_v>0$~\cite{SM}. The energies enter only through powers whose moduli are constant along these contours, so convergence does not depend on the kinematics. A vertex with no external leg, such as the central vertex of the triple exchange, has $x_v=0$. Its time integral is instead a delta function in the Mellin variables~\cite{SleightTaronna2021b}, which sets $r_v+\Sigma_v=0$ and removes one contour with weight $\half$. This rule holds for $0<r_v<\deg(v)$ when the neighboring vertices carry external legs~\cite{SM}. We evaluate Eq.~(\ref{eq:tree}) by quadrature along the contours, or close them on the evenly spaced poles of its Gamma functions.

Equal-time correlators follow from the same time integrals by cutting internal lines into blocks~\cite{Chen2017SK,Tong2022,SM}. A cut line, with its ends on opposite branches of the in-in contour, carries $\frac1\pi(t\,t')^{3/2}K_\nu(yt)K_\nu(yt')$. An uncut line carries $(t\,t')^{3/2}$ times the Feynman kernel $G^F_\nu=\bigl[e^{\pi\mu}\G_\nu-e^{-\pi\mu}\G_{-\nu}\bigr]/(2\sinh\pi\mu)$, the Bunch--Davies combination of Refs.~\cite{SleightTaronna2021a,SleightTaronna2021b}, each term represented by Eq.~(\ref{eq:kernel}). Every block is again of the form of Eq.~(\ref{eq:tree})~\cite{SM}. The correlator is the sum over subsets of $n_c$ cut lines of $(-1)^{L-n_c}(2/\pi^{n_c})$ times the real part of the product of the blocks~\cite{SM}. When every vertex phase is real, as for $\pid$ legs, the blocks on either side of each cut line are conjugated relative to each other (otherwise each block keeps its own phase).

\begin{figure}[t]
\centering
\includegraphics[width=\columnwidth]{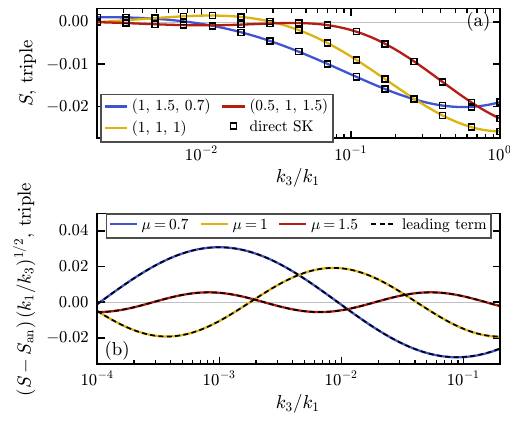}
\caption{Triple-exchange bispectrum at physical kinematics, $S=(k_1k_2k_3)^2\,\langle\pi\pi\pi\rangle'$ for one leg assignment, at unit couplings and $k_1=k_2$. (a)~Curves, from Eq.~(\ref{eq:tree}) or~(\ref{eq:master}), labeled by the mass parameters $\mu_l=(m_l^2-\frac94)^{1/2}$ of the lines $A,B,C$, with the line on the $\bk_3$ vertex last. Squares are direct Schwinger--Keldysh integration~\cite{SM}. (b)~Squeezed collider signal of (a), $(S-\San)(k_1/k_3)^{1/2}$ with the fitted polynomial background $\San$ subtracted (\EM{D}), for the three soft masses $\mu$ (color), and the leading term $2\Re[\cM_3(k_3/k_1)^{\ii\mu}]$ of Eq.~(\ref{eq:M3}) (dashed).}
\label{fig:trees}
\end{figure}

% ===========================================================================
\textit{Triple exchange at physical kinematics.}---As the test on trees we take the triple exchange [Fig.~\ref{fig:diagrams}(a)] at leading order in the mixing, in which each of three principal-series fields $\sigma_l$, $l=A,B,C$, couples to the inflaton through a mixing vertex $\rho_l\,\pid\sigma_l$, and the three meet at a cubic vertex $g\,\sigma_A\sigma_B\sigma_C$ with no external leg. We evaluate it from Eq.~(\ref{eq:tree}) with the rule for that vertex and assemble it with the cutting rule on every line. Figure~\ref{fig:trees}(a) follows isosceles triangles from equilateral down to $k_3/k_1=2\times10^{-3}$. Throughout that range the bispectrum agrees pointwise with direct Schwinger--Keldysh integration~\cite{Chen2017SK}, and the double- and triple-exchange correlators agree with the transport code of Ref.~\cite{Werth2024} to a median 0.1\% over 48 configurations, scalene triangles included~\cite{SM}. The double exchange, whose cubic vertex $\pid\sigma_1\sigma_2$ carries an external leg, reproduces the known series solution~\cite{Aoki2024,SM}.

Closing first the contours at the mixing vertices, where the energies of the line and the leg coincide, turns each line into a dressed leg (the mixed propagator of Ref.~\cite{Chen2017SK}), a Laplace transform of Legendre functions~\cite{Pinol2026strong,Belrhali2026laplace,Belrhali2026strong} with density $f_l$ (\EM{B}). The central time integral is then elementary, and for one leg assignment
\begin{equation}
S=-\frac{k_1k_2k_3}{2}\,\Re\!\int_1^\infty\!\!\dd^3z\;\frac{F_A(z_1)\,F_B(z_2)\,F_C(z_3)}{(k_1z_1+k_2z_2+k_3z_3)^3}\,, \label{eq:master}
\end{equation}
with $F_l(z)=f_l\bigl(\frac{z-1}{2}\bigr)$, so that the masses sit in the densities and the momenta in the kernel. It has the form of the single integral over dressed legs of Ref.~\cite{Pinol2026strong}, here with three different masses, and for one heavy field at any mixing Ref.~\cite{ChenXianyu2026} expands the same integral in a convergent double series. Direct quadrature of Eq.~(\ref{eq:master}) reproduces Fig.~\ref{fig:trees}(a) to $5\times10^{-10}$. Expanding only the leg with the smallest momentum, with the other two kept exact, gives a series in $u_3=2k_3/k_T\le\frac23$, $k_T=k_1+k_2+k_3$, whose terms are sums of hypergeometric functions (\EM{B}). Its expansion parameter never exceeds $u_3$, so it converges at every physical shape, folded ones included. Ref.~\cite{Aoki2026triple} expands the same leg. Expanding all three legs instead gives the series of Ref.~\cite{XianyuZang2026}, which converges only for $u_1+u_2+u_3<1$, below the physical value 2~\cite{SM}.

In the squeezed limit $k_3\ll k_1\simeq k_2$, with $\bk_3$ on a mixing vertex [Fig.~\ref{fig:diagrams}(a)], the shape oscillates at the adjacent line's mass alone~\cite{Chen2017SK}, $S\to\San+2\Re[\cM_3\,(k_3/k_1)^{1/2+\ii\mu_C}]$ with $\San$ analytic. The soft line supplies one Boltzmann factor $e^{-\pi\mu_C}$ and the thermal weights of the cut, and the rest of the diagram supplies time integrals that carry no Boltzmann factor of the hard lines' masses. At general masses $\cM_3$ is a single sum of products of ${}_3F_2$ at unit argument (\EM{B}). At $\mu=1$ on every line and unit couplings, the double- and triple-exchange coefficients $|\cM_2|$ and $|\cM_3|$ are $0.37$ and $0.32$ times the single-exchange coefficient~\cite{SM}, whereas a second Boltzmann factor would have cost $e^{-\pi}\approx0.04$. This confirms at general masses, and to all orders in the Boltzmann factor, that multi-exchange signals are not exponentially smaller than single exchange~\cite{Lee2016,Kumar2018,Kumar2026planck,Tong2022,Aoki2024,Chen2017SK,XianyuZang2026}. At equal masses $\cM_3$ agrees with the closed form of Ref.~\cite{Pinol2026strong} (\EM{B}) and with the coefficient of Ref.~\cite{XianyuZang2026}~\cite{SM}.

% ===========================================================================
\textit{One loop.}---At fixed loop momentum $\bl$ every line of a one-loop diagram has a definite energy $y_l(\bl)$, and its time integrals factorize as for a tree, so Eq.~(\ref{eq:tree}) gives them unchanged~\cite{SleightTaronna2021b}. Each line contributes the power $(y_l/2)^{-2\delta_l}$, with $\delta_l$ fixed at each point of the contours, so we can do the momentum integral first. For two lines with energies $y_1=|\bl|$ and $y_2=|\bk-\bl|$ it is the flat-space two-line integral with arbitrary powers~\cite{BoosDavydychev1991,QinXianyu2023loop},
\begin{align}
\int\!\dd^3\ell\;\, y_1^{-2\delta_1}\,y_2^{-2\delta_2} &= k^{3-2\cS}\,C_2(\delta_1,\delta_2)\,, \nonumber\\
C_2 &= \pi^{3/2}\,\frac{\Gamma\bigl(\thalf-\delta_1\bigr)\Gamma\bigl(\thalf-\delta_2\bigr)}{\Gamma(\delta_1)\,\Gamma(\delta_2)}\,\frac{\Gamma\bigl(\cS-\thalf\bigr)}{\Gamma(3-\cS)}\,, \label{eq:C2}
\end{align}
with $k=|\bk|$ and $\cS=\delta_1+\delta_2$. The momentum integral converges only for $\Re\delta_1,\Re\delta_2<\thalf<\Re\cS$. The Mellin--Barnes integral with $C_2$ inserted converges in the bubble only for $\Re\cS<\thalf$, because along one direction of the contours the integrand decays only as a power (\EM{C}). Commuting the two integrals therefore crosses the pole of $\Gamma(\cS-\thalf)$~\cite{Premkumar2024}, whose residue carries the logarithm of the loop.

We take the bubble with two conformally coupled legs at each vertex ($r_v=2$), vertex energies $x_1$, $x_2$, and line masses $m_1$, $m_2$. Cut off at comoving $|\bl|=\Lambda$, the loop integral is a four-variable Mellin--Barnes integral, $\psi_2^{\rm MB}$, on contours with $\half<\Re\cS<\thalf$, plus $\pi\ln\Lambda/(x_1+x_2)$ and a constant independent of $k$, both from the residue of the pole crossed~\cite{SM}. The divergence thus has a mass-independent coefficient and the energy dependence of a contact term, removed by one quartic counterterm of the external legs. Integrating first over the two Mellin directions that the energies do not enter gives the bubble at unequal masses as a sum of single exchanges,
\begin{equation}
\psi_2^{\rm MB}=8\pi\sum_{n\ge0}G_n\bigl[\psi_1^{(\nu'_n)}(k)-\psi_1^{(\nu'_n)}(0)\bigr]\,, \label{eq:tower}
\end{equation}
with $\nu'_n=\nu_1+\nu_2+\thalf+2n$. Here $\psi_1^{(\nu)}(k)$ is the single exchange of order $\nu$ through momentum $k$ at vertex powers $r_v-\thalf$, its value at $k=0$ is elementary, and $G_n$ is a ratio of Gamma functions (\EM{C}). The terms fall as $n^{-3}$, and the sum converges at every shape short of the folded limit~\cite{SM}. This is the Mellin form of the product identity of bulk-to-bulk propagators~\cite{FitzpatrickKaplan2012a,FitzpatrickKaplan2012b,SleightTaronna2021b}, and the same structure appears at equal masses in Ref.~\cite{Liu2026spectral}. At equal masses the residue series match the spectral decomposition of Ref.~\cite{XianyuZhang2023} term by term, and at unequal masses the assembled bispectrum matches direct time integration to $3\times10^{-12}$~\cite{SM}.

% ===========================================================================
\textit{The triangle.}---In the triangle of Fig.~\ref{fig:diagrams}(b) the line energies are $y_A=|\bl+\bk_1|$, $y_B=|\bl-\bk_3|$, and $y_C=|\bl|$. The momentum factor is the flat-space three-line integral with arbitrary powers, $C_3(\delta;k)=\int\dd^3\ell\,\prod_l(y_l^2)^{-\delta_l}$~\cite{QinXianyu2023loop,BoosDavydychev1991,Davydychev1992}, which we use in a Mellin--Barnes form valid at physical momenta (\EM{C}). The triangle is then Eq.~(\ref{eq:tree}) with $\prod_l(y_l/2)^{-2\delta_l}$ replaced by $4^{\cS}C_3$, with $\cS=\delta_A+\delta_B+\delta_C$. With one $\pid$ leg at each vertex ($r_v=2$), the case entering the inflaton bispectrum, the vertex conditions and the decay of the integrand allow $\Re\cS<\frac52$. The momentum and contour integrals then commute, and the triangle is a finite integral with no logarithm (\EM{C}). On straight contours at $\Re\delta_l=0.6$ it matches brute-force integration at 14 configurations to a median $5.5\times10^{-6}$, and the brute-force integration agrees with Ref.~\cite{Herderschee2026} at equal masses to 0.2\%~\cite{SM}. Our triangle results were obtained independently of Refs.~\cite{Herderschee2026,QinXianyu2026triangle}.

With the soft leg on the vertex joining lines $B$ and $C$, at $k_1=k_2=k$, the non-analytic terms come from soft loop momenta. We insert an expansion of $C_3$ in $k_3^2/k^2$ that converges at every physical shape~\cite{SM} and close the contours of the two soft lines. Two first Barnes lemmas and a selection rule then reduce each order to a finite sum, and the non-analytic part of the bispectrum is
\begin{multline}
S\supset2\Re\sum_{N\ge0}\Big(\frac{k_3}{k}\Big)^{2+2N}\Big[\cM^{(N)}_{+}\Big(\frac{k_3}{k}\Big)^{\ii(\mu_B+\mu_C)}\\
+\cM^{(N)}_{-}\Big(\frac{k_3}{k}\Big)^{\ii(\mu_B-\mu_C)}\Big], \label{eq:Ssq}
\end{multline}
with each $\cM^{(N)}_\pm$ a finite double sum of Gamma functions and ${}_3F_2$ at unit argument (\EM{C}). The first family is the pair oscillation, with exponent 2 in place of $\half$ and one Boltzmann factor per produced particle, $\cM^{(0)}_+\propto e^{-\pi(\mu_B+\mu_C)}$. At unequal masses the second family oscillates at the difference frequency, as in the fermion bubble~\cite{Aoki2026fermion}, with the same Boltzmann factor and a size comparable to the pair family. At $\mu_B=\mu_C$ it becomes analytic in $k_3^2$, with no logarithm~\cite{SM}. At $N=0$ the pair coefficient reduces at equal masses to that of Ref.~\cite{QinXianyu2026triangle}, and brute-force integration of the assembled correlator reproduces both families to $7\times10^{-5}$. The series converges for $k_3<2k$, and the first correction is about half the leading term at $k_3=k$ at every mass we tried~\cite{SM}. A free fit to direct time integration returns $\cM^{(1)}_+$ within 3\% of its closed value~\cite{SM}.

With conformally coupled legs ($r_v=1$) the two convergence regions only touch, and continuing across the pole of $\Gamma(\cS-\thalf)$ leaves a logarithm $4\pi D_0\ln\Lambda$ whose coefficient $D_0$ depends on the masses and not on the momenta (\EM{C}). The logarithm is secular, not ultraviolet, and conformal symmetry fixes $\psi_3$ to $-4\pi D_0\ln k_T+b$~\cite{Bzowski2016}, which our evaluation confirms to $2\times10^{-6}$~\cite{SM}. In the equal-time correlator the logarithm cancels, as it must for operators of dimension one, which we confirm with closed forms to $10^{-30}$~\cite{SM}. What remains is one mass-dependent constant times $1/k_1k_2k_3$ (\EM{C}). The selection rule removes every soft-limit oscillation of this triangle at every order, as the conformal form of $\psi_3$ requires, so in the triangle oscillatory signals require the boost-breaking $\pid$ legs. When the lines are conformally coupled as well, with two legs at each vertex, the triangle correlator is known in closed form~\cite{Chowdhury2026,Pimentel2026,Henn2026}.

% ===========================================================================
\textit{Strong mixing.}---At $\rho\gtrsim1$ the mixing $\rho\,\pid\sigma$ must be resummed, and the oscillation frequency is set by $\mu_{\rm eff}^2=m^2+\rho^2-\frac94$~\cite{An2018nonpert,Iyer2018strong,Wang2026strong,Huenupi2026,Pinol2026,Belrhali2026strong}. We take the cubic coupling $-\frac\alpha2\pid\sigma^2$ of Refs.~\cite{Pinol2026strong,Belrhali2026strong}, whose one-loop bispectrum contains the triangle at equal masses with dressed lines and legs. The modes in the Laplace representation of Ref.~\cite{Belrhali2026strong} turn the soft lines and their shared vertex into a two-fold Mellin--Barnes integral with $C_2$ of Eq.~(\ref{eq:C2}), and the hard line is the time-ordered dressed propagator~\cite{SM}. Because the coupling contains $\pid$, the interaction Hamiltonian also carries a contact term $\frac{\alpha^2}{8}a^3\sigma^4$~\cite{Abolhasani2022,Pinol2026strong}, which along $m=0$ is 0.7 to 1.5 times the triangle. At $\rho=0.01$ the sum reproduces $\cM^{(0)}_+$ of Eq.~(\ref{eq:Ssq}) to $5\times10^{-4}$ at eight masses. Along $m=0$, the largest mixing without a gradient instability~\cite{SM}, the pair coefficient falls as $e^{-\pi\mu_{\rm eff}}$ up to a power. Each produced particle then pays half the Boltzmann exponent, as at tree level~\cite{Belrhali2026strong,Pinol2026strong}, while at fixed $\rho$ the coefficient keeps $e^{-2\pi\mu_{\rm eff}}$~\cite[Fig.~S5]{SM}. The tree-level signal of the same coupling is far larger, however~\cite[Table~SII]{SM}. At $k_3/k=0.1$ the loop's oscillation is of order $10^{-4}$ of the tree's at $\mu_{\rm eff}=1$ and $10^{-8}$ at $\mu_{\rm eff}=5$ along $m=0$. It stays below $10^{-3}$ for $1\le\mu_{\rm eff}\le5$, even with $\alpha$ at the perturbativity bound $(\rho/H)^{3/4}$ of Ref.~\cite{Philcox2026lim}. Loops need not be a correction, though. If a field that cannot mix runs in the loop while the inflaton mixes strongly with a singlet, its pair signal has no tree-level counterpart. The reduction covers this case, with dressed legs and undressed lines, though we have not computed it.

% ===========================================================================
\textit{Discussion.}---The leading squeezed term describes a loop signal only deep in the squeezed limit, and at unequal masses the triangle carries a second, slower oscillation of comparable size. In the triangle, oscillatory signals require the boost-breaking $\pid$ couplings. At weak mixing, each line carrying the soft momentum in a squeezed limit pays one Boltzmann factor, while each hard line costs only an $O(1)$ factor at masses of order $H$~\cite{Lee2016,Tong2022,Aoki2024,XianyuZang2026}. At strong mixing, loops can be left out of templates in the model considered, though not when the field in the loop cannot mix.

The reduction behind these results applies to massive diagrams in de Sitter space generally, and Eq.~(\ref{eq:tree}) evaluates the triple exchange at general kinematics in 0.1\,s per leg assignment, fast enough for templates~\cite{SM}. Spinning lines change the vertex powers and add a polynomial polarization sum, and we expect the pole analysis to carry over to the box and to two-loop diagrams.

\textit{Acknowledgments.}---The author thanks Oliver Philcox, Soubhik Kumar, David Dunsky, and Graham van Goffrier for comments on the manuscript, and Lucas Pinol for helpful correspondence. This work was carried out under the Anthropic STEM Fellows Program. Code development, calculations, and manuscript preparation were assisted by Claude Fable 5.1, with further checks by Claude Fable 5.2 and Claude Opus 5.5. The author verified the results and takes full responsibility for them. The views expressed are the author's own and are not endorsed by Anthropic.

% ===========================================================================
% End Matter
% ===========================================================================
\section*{End Matter}

\phantomsection\label{app:A}
\textit{A. Representation and conventions.}---Equation~(\ref{eq:kernel}) is the propagator representation of Refs.~\cite{SleightTaronna2021a,SleightTaronna2021b} (see also Sec.~4.7 of Ref.~\cite{SleightTaronna2020}), stripped of $(tt')^{3/2}$, in the variables $s=-u-\nu/2$, $s'=-\bar u-\nu/2$, with the same integrand and normalization. It also follows from the Bessel product by Barnes' method~\cite{Barnes1908,Watson1944}. Its pair $\Gamma(\delta)\Gamma(1-\delta)$ is the cosecant contact-term factor of Refs.~\cite{Sleight2020,SleightTaronna2020,SleightTaronna2021a,SleightTaronna2021b}, whose split into two pole families is the minimal choice of improvement terms of Ref.~\cite{SleightTaronna2020}, and the strip $0<\Re\delta<1$ appears in Ref.~\cite{SleightTaronna2021b}. The vertex rule is that of the partial Mellin--Barnes method~\cite{XianyuZang2024,QinXianyu2022phase,QinXianyu2023helical}, and holds unchanged at vertices joined by time-ordered lines because Eq.~(\ref{eq:kernel}) stores the ordering. It also follows from the Mellin--Barnes Feynman rules of Ref.~\cite{SleightTaronna2021b}, in which each vertex carries a Dirac delta in the Mellin variables, with the lifting formula of Ref.~\cite{Sleight2020} for conformally coupled legs. A vertex without legs keeps the delta, with weight $\half$ because $\partial\Sigma_v/\partial s=2$. New here are the absolute convergence of Eq.~(\ref{eq:kernel}) for $0<\Re\delta<\half$ and of Eq.~(\ref{eq:tree}) for $0<\Re\delta_l<1$, the existence of straight contours whenever every $r_v>0$, and the sufficient condition $0<r_v<\deg(v)$ for the delta rule when the neighboring vertices carry external legs~\cite{SM}. Vertices with $r_v\le0$ are defined by continuation in $r_v$. From the coefficient of Ref.~\cite{ArkaniHamed2017} we strip the couplings $g_v$, one factor $\ii/\eta_0$ per conformally coupled leg, and a factor $-1$ per internal line, so that $\psi^{\rm full}=(-1)^L\,\ii^N\eta_0^{-N}\bigl(\prod_v g_v\bigr)\psi_n$ for $L$ internal lines and $N$ such legs. Every coupling enters $\Lint$ with a plus sign. The kernel differs from the Dirichlet propagator of Ref.~\cite{ArkaniHamed2017} by a boundary term that cancels in correlators~\cite{SM}.

\phantomsection\label{app:B}
\textit{B. Series and squeezed coefficient of the triple exchange.}---At a mixing vertex $x_l=y_l=k_l$, so the $s_l$ integral is the mixing time integral in closed form. In the correlator it gives the dressed leg $w_l(k_lt_v)$, which following Refs.~\cite{Pinol2026strong,Belrhali2026laplace,Belrhali2026strong} we write as
\begin{align}
w_l(t)&=2t^{1/2}e^{-t}\!\int_0^\infty\!\!\dd x\,e^{-2tx}f_l(x)\,, \nonumber\\
f_l(x)&=Q_{d_l-1}(1+2x)+c^+_lP_{d_l-1}(1+2x)\,, \label{eq:leg}
\end{align}
with $d_l=\half-\ii\mu_l$ and $c^\pm_l=-\frac\pi2(1\pm\ii e^{\pm\pi\mu_l})/\cosh\pi\mu_l$, and $P_{d-1}$ and $Q_{d-1}$ are the Legendre functions of argument above one~\cite{SM}. The central time integral then gives Eq.~(\ref{eq:master}). With $u_l=2k_l/k_T$, $z_l=1+2x_l$, and $D=1+u_1x_1+u_2x_2\ge1$, the leg on $\bk_3$ enters through $D^{-3}\!\int_0^\infty\dd x\,f_C(x)(1+u_3x/D)^{-3}$, one ${}_3F_2$ and two ${}_2F_1$ in $u_3/D$. Their expansion $\sum_ea_e(u_3/D)^e$ runs over the exponents $e=n$, $d_C-1+m$, and $m-d_C$ ($n,m\ge0$), the analytic background and the two signal families~\cite{SM}. In $t_i=u_ix_i/(1+u_ix_i)$ the other two legs factorize, since $D=(1-t_1t_2)/[(1-t_1)(1-t_2)]$, and expanding $(1-t_1t_2)^{-3-e}$ gives
\begin{align}
S&=-\frac{u_3}{2}\,\Re\sum_ea_eu_3^{\,e}\,\mathcal{J}(3{+}e)\,, \nonumber\\
\mathcal{J}(s)&=\sum_{k\ge0}\frac{(s)_k}{k!}\,W^A_k\bigl[\tfrac1{u_1}\bigr](s)\,W^B_k\bigl[\tfrac1{u_2}\bigr](s)\,,\nonumber\\
W_k[c](s)&=\int_0^1\!\dd t\;t^k(1-t)^{s-2}f\Bigl(\frac{ct}{1-t}\Bigr)\,, \label{eq:triple}
\end{align}
each $W_k$ a single sum of ${}_2F_1$~\cite{SM}. With $\bk_3$ the smallest momentum, $u_3/D\le u_3\le\frac23$, inside the unit radius of convergence of this expansion~\cite{SM}, so Eq.~(\ref{eq:triple}) converges at every physical shape. It reproduces Eq.~(\ref{eq:tree}) to $3\times10^{-8}$ or better from $(1,1,0.02)$ to the folded $(1,0.5,0.5)$~\cite{SM}. For one heavy field at any mixing, Ref.~\cite{ChenXianyu2026} obtains convergence at physical kinematics through an order of summation, while here it comes from keeping the hard legs exact and expanding only the softest leg. The two series were obtained independently. Ref.~\cite{Aoki2026triple} expands the same mixed propagator and obtains the dependence on the ratio of the hard momenta from a differential equation. Its squeezed coefficient at unequal masses is likewise a single sum of products of ${}_3F_2$ at unit argument.

At $u_1=u_2=1$ the moment of a density $Q_{d-1}+c\,P_{d-1}$ is $W_k=Q_k+c\,P_k$, with
\begin{align}
Q_k&=\frac{\Gamma(d)^2\,(k!)^2}{2\,\Gamma(k{+}1{+}d)^2}\;{}_3F_2\!\left(\begin{matrix}1{+}d{-}s,\,k{+}1,\,k{+}1\\ k{+}1{+}d,\,k{+}1{+}d\end{matrix};1\right), \nonumber\\
P_k&=B(k{+}1,s{+}d{-}1)\;{}_3F_2\!\left(\begin{matrix}d,\,d,\,k{+}1\\ 1,\,s{+}d{+}k\end{matrix};1\right). \label{eq:QP}
\end{align}
With $\bar s=\frac52-\ii\mu_C$, the thermal cut weights $\lambda_\pm=\pm(e^{\pm\pi\mu_C}\mp\ii)/(2\sinh\pi\mu_C)$, and the unit-energy mixing block $\hat\Phi_C=\sqrt{\pi/2}\,\Gamma(\half{+}\ii\mu_C)\Gamma(\half{-}\ii\mu_C)$, the squeezed coefficient is
\begin{align}
\cM_3&=-\frac{\hat\Phi_C\,2^{-1/2-2\ii\mu_C}}{8\,\Gamma(1+\ii\mu_C)}\,\overline{\bigl[\bar\lambda_+\mathcal{J}_1(\bar s;c^-)+\bar\lambda_-\mathcal{J}_1(\bar s;c^+)\bigr]}\,, \nonumber\\
\mathcal{J}_1(s;c)&=\Gamma(s)\sum_{k\ge0}\frac{(s)_k}{k!}\,\bigl(Q^A_k+c_AP^A_k\bigr)\bigl(Q^B_k+c_BP^B_k\bigr)\,, \label{eq:M3}
\end{align}
with $c^\pm=(c^\pm_A,c^\pm_B)$. At equal masses it agrees with the closed form of Ref.~\cite{Pinol2026strong}, $A_3e^{\ii\delta_3}/\lambda^3=64\ii\,\overline{\cM_3}$ in the notation of its Eq.~(6.17), to $4\times10^{-10}$ or better at seven masses, $0.5\le\mu\le4$.

\phantomsection\label{app:C}
\textit{C. Loop integrals and the triangle.}---Loop integrals in the coefficients $\psi$ use the bare measure $\dd^3\ell$, and correlators include the physical $(2\pi)^{-3}$. Equation~(\ref{eq:C2}) follows with Schwinger parameters. Along one direction of the contours, which shifts the Mellin variables by $\pm\ii\lambda$ alternately around the loop so that every line and vertex sum stays fixed, the integrand is not exponentially suppressed. In the bubble it decays only as $|\Im s|^{2\cS-4}$, which requires $\Re\cS<\thalf$, and at $\cS=\thalf$ this tail diverges. Across the poles of $\Gamma(\cS-\thalf)$ the loop integral jumps by their residues, so $\Re\cS>\half$ is part of its definition. The weights of Eq.~(\ref{eq:tower}) are, with $\bar\nu=\nu_1+\nu_2$,
\begin{equation}
G_n=\frac{\thalf+\bar\nu+2n}{(1+\bar\nu+2n)(2+\bar\nu+2n)}\prod_{a=0,\nu_1,\nu_2,\bar\nu}\frac{\Gamma(\thalf+a+n)}{\Gamma(1+a+n)}\,, \label{eq:Gn}
\end{equation}
and the right side of Eq.~(\ref{eq:tower}) is the Mellin form of the product identity, each term the integrand of a single exchange~\cite{SM}. For three lines the Mellin--Barnes splitting of $(1+a+b)^{-c}$ gives the two-variable form of Refs.~\cite{QinXianyu2023loop,BoosDavydychev1991}. With $z=k_1^2/k_3^2$, $w=k_2^2/k_3^2$, and $\cN=\pi^{3/2}(k_3^2)^{\frac32-\cS}/[\Gamma(\delta_A)\Gamma(\delta_B)\Gamma(\delta_C)\Gamma(3-\cS)]$,
\begin{align}
C_3 &= \cN\iint\!\frac{\dd u\,\dd u'}{(2\pi\ii)^2}\,\Gamma(-u)\Gamma(-u')\,z^uw^{u'}\,\Gamma(\cS{-}\thalf{+}u{+}u') \nonumber\\
&\qquad\times\Gamma(\delta_A{+}u{+}u')\,\Gamma(\thalf{-}\delta_A{-}\delta_C{-}u) \nonumber\\
&\qquad\times\Gamma(\thalf{-}\delta_A{-}\delta_B{-}u')\,, \label{eq:C3}
\end{align}
on contours separating the poles of $\Gamma(\cS{-}\thalf{+}u{+}u')$ and $\Gamma(\delta_A{+}u{+}u')$ from those of the other four Gamma functions. Unlike its Appell $F_4$ series~\cite{BoosDavydychev1991,Davydychev1992}, Eq.~(\ref{eq:C3}) holds at physical momenta~\cite{SM}.

In the triangle at fixed $\bl$ every line end meets one vertex, so $\sum_v(r_v+\Sigma_v)=\sum_vr_v-2\cS$. For $r_v=2$ the right side is $6-2\cS$, and with the decay $|\Im s|^{2\cS-6}$ along the direction above the contours reach $\Re\cS<\frac52$, overlapping the domain $\Re\cS>\thalf$ of the loop integral. For $r_v=1$ it is $3-2\cS$, which confines the contours to $\Re\cS<\thalf$. The two regions then touch at the pole of $\Gamma(\cS-\thalf)$, which is pinched. Regulating the measure as $\dd^3\ell\,\ell^{-2\epsilon}$ opens a window between them, and deforming back to $\Re\cS>\half$ crosses only that pole, whose residue $2\pi$ is independent of masses and momenta. The integrand left on that pole has a simple pole of its own at $\cS=\thalf$, where the three vertex residues strip every power of $x_v$. It gives the logarithm $4\pi D_0\ln\Lambda$, with a coefficient independent of the momenta, and a finite remainder~\cite{SM}, where
\begin{equation}
D_0 = \int_0^\infty\prod_{v=1}^3\dd t_v\prod_l\G_{\nu_l}(1;\,t_{v_l},t_{v_l'})\,, \label{eq:D0}
\end{equation}
equal to $\frac18\,\zeta(3,\half+\ii\mu)$ at equal masses, with $\zeta(s,a)$ the Hurwitz zeta function. In the correlator, writing every kernel through $\G_{\pm\nu}$ makes the coefficient of the logarithm proportional to $\Im Q$, with $Q$ a sum of Eq.~(\ref{eq:D0}) at the eight sign-flipped orders with real weights. With the closed form of $D_0$ at general masses~\cite{SM}, $|\Im Q|/|Q|$ is below $10^{-30}$. The correlator is then proportional to $\eta_0^3\,\mathfrak{B}/k_1k_2k_3$, with $\mathfrak{B}=4\pi^2Q+\Delta$. The first term comes from the Lorentzian phase of the late-time logarithm, and $\Delta$ is Boltzmann suppressed. At $\mu=(1,1,1)$, $\mathfrak{B}=-3.11475$~\cite{SM}.

For the triangle with $\pid$ legs, squeezed on the vertex joining $B$ and $C$ at $k_1=k_2=k$, the coefficient is $\psi_3=\sum_Na_Nk_3^N+k^{-3}(k_3/k)^{1+\nu_B+\nu_C}\sum_Nc_N(k_3/k)^{2N}$, with no other non-analytic family. The pair family comes from the poles of $\Gamma(-s)$ at the far ends of the two soft lines, and
\begin{multline}
c_N=\frac{\pi^2}{2^{1+\nu_{BC}}}\,\frac{\Gamma(-1-\nu_{BC}-2N)}{2+\nu_{BC}+2N}\sum_{i+j+n=N}\frac{4^{-i-j}}{i!\,j!\,n!}\\
\times\frac{\Gamma(\frac32{+}\nu_B{+}2j{+}n)\,\Gamma(\frac32{+}\nu_C{+}2i{+}n)}{\Gamma(1{+}\nu_B{+}j)\,\Gamma(1{+}\nu_C{+}i)}\\
\times2\big(-\tfrac12y\partial_y\big)_n\mathcal{H}_{ij}(y)\Big|_{y=1}, \label{eq:cN}
\end{multline}
with $\nu_{BC}=\nu_B+\nu_C$, $(D)_n=D(D+1)\cdots(D+n-1)$, and the hard factor
\begin{equation}
\mathcal{H}_{ij}(y)=\!\int_0^\infty\!\!\dd t_1\dd t_2\,t_1^{1+\nu_C+2i}t_2^{1+\nu_B+2j}e^{-t_1-t_2}\G_{\nu_A}(y;t_1,t_2). \label{eq:Hij}
\end{equation}
At $i=j=0$ this is the doubly folded limit of the single-exchange seed integral of Ref.~\cite{QinXianyu2023closed} [its Eqs.~(92)--(93)], and its derivatives at $y=1$ are the same seed at integer-shifted exponents, each a ratio of Gamma functions times one ${}_3F_2$ at unit argument. The selection rule comes from a factor $1/\Gamma(r_3-1-J-n+K+q)$ left by the second of the two Barnes lemmas~\cite{SM}. Here $r_3$ is the power at the soft vertex, and the integers $K\le J=i+j$ and $q\le n$ label the terms of two polynomial expansions. At $r_3=2$ only $K=J$ and $q=n$ survive, and at $r_3=1$ none does. Each $\cM^{(N)}_\pm$ of Eq.~(\ref{eq:Ssq}) is a sum of $c_N$ at four sign-flipped orders with real weights~\cite[Eq.~(S55)]{SM}. When the soft line at a vertex is degenerate with the hard line, $\cM^{(0)}_+$ is a ratio of Gamma functions, and for three equal masses it coincides with Eq.~(26) of Ref.~\cite{QinXianyu2026triangle} to $10^{-16}$.

\phantomsection\label{app:D}
\textit{D. Numerical checks.}---The tree-level results and the loop evaluations of Figs.~S3 and S4 use the code of Ref.~\cite{McCuneCode}, and the closed forms and strong-mixing results use further scripts by the author~\cite{SM}. Contour integrals use trapezoidal quadrature on straight contours spot-checked at 25 digits, and we quote relative differences. Direct Schwinger--Keldysh time integration reproduces the double and triple exchange to $2\times10^{-10}$ or better at 198 isosceles and scalene configurations down to $k_3/k_1=0.002$. In Fig.~\ref{fig:trees}(b) $\San$ is the degree-six polynomial of a fixed-exponent fit over $10^{-4}\le k_3/k_1\le0.25$.

% ===========================================================================
% Supplemental Material
% ===========================================================================
\clearpage
\onecolumngrid
\begin{center}
{\large\bfseries Supplemental Material}
\end{center}
\vspace{1em}
\setcounter{equation}{0}\setcounter{figure}{0}\setcounter{section}{0}\setcounter{secnumdepth}{1}
\renewcommand{\theequation}{S\arabic{equation}}
\renewcommand{\thefigure}{S\arabic{figure}}
\renewcommand{\thetable}{S\Roman{table}}
\renewcommand{\thesection}{S\arabic{section}}
\renewcommand{\theHequation}{S\arabic{equation}}
\renewcommand{\theHfigure}{S\arabic{figure}}
\renewcommand{\theHsection}{S\arabic{section}}

Equation, figure, and section numbers without a prefix S refer to the main text, and reference numbers to the reference list of the main text.

\section{The kernel identity and its convergence window}\label{sm:kernel}

\textit{Ingredients.}---We use the series of $I_\nu$ and the Barnes integral of $K_\nu$,
\begin{align}
I_\nu(z)&=\sum_{m\ge0}\frac{(z/2)^{\nu+2m}}{m!\,\Gamma(\nu+m+1)}\,, \nonumber\\
K_\nu(z)&=\frac12\int\!\frac{\dd s'}{2\pi\ii}\,\Gamma(-s')\Gamma(-\nu-s')\Bigl(\frac z2\Bigr)^{\nu+2s'}, \label{sm:eq1}
\end{align}
the latter on a line $\Re s'<\min(0,-\Re\nu)$ with both pole series to its right. We write the right side of Eq.~(2) as $\mathcal R(t,t')$, the integral of $\frac12f(s,s')\,t^{\nu+2s}t'^{\,\nu+2s'}$ over $\dd s\,\dd s'/(2\pi\ii)^2$, with
\begin{equation}
f(s,s')=\frac{\Gamma(-s)\Gamma(-s')\Gamma(\delta)\Gamma(1-\delta)}{\Gamma(\nu+s+1)\Gamma(\nu+s'+1)}\Bigl(\frac y2\Bigr)^{-2\delta}, \label{sm:eq2}
\end{equation}
$\delta=-\nu-s-s'$. We take $\Re\nu\ge0$, which covers both series and the reflected order $-\ii\mu$. At fixed $s'$ the poles of $\Gamma(-s)$ at $s=m$ and of $\Gamma(\delta)$ at $s=n-\nu-s'$ ($m,n\ge0$) lie to the right. Those of $\Gamma(1-\delta)$ at $\delta=1+n$ lie to the left, and likewise in $s'$. Straight lines separate the two sets exactly on the window
\begin{equation}
\Re s<0\,,\qquad \Re s'<0\,,\qquad 0<\Re\delta<1\,, \label{sm:eq3}
\end{equation}
where $f$ is analytic. Hence all placements of the lines give the same $\mathcal R$ wherever the integral converges.

\textit{Absolute convergence.}---Write $s=\sigma+\ii\tau$, $s'=\sigma'+\ii\tau'$. By Stirling's formula the exponentials of $\Gamma(-s)$ and $1/\Gamma(\nu+s+1)$ cancel at large $|\tau|$, leaving $|\tau|^{-2\sigma-1-\Re\nu}$, and similarly for $s'$. Further, $|\Gamma(\delta)\Gamma(1-\delta)|=\pi/|\sin\pi\delta|\sim2\pi e^{-\pi|\tau+\tau'+\Im\nu|}$. The times enter through $|t^{2\ii\tau}t'^{\,2\ii\tau'}|=1$. In $u=\tau+\tau'$, $w=\tau-\tau'$ the integrand is therefore bounded by a polynomial in $|u|$ times
\begin{equation}
e^{-\pi|u|}\,(1+|w|)^{2\Re\delta-2}\,, \label{sm:eq4}
\end{equation}
since $-2(\sigma+\sigma')-2-2\Re\nu=2\Re\delta-2$.
The decay is exponential along the diagonal but only a power law along the anti-diagonal. The double integral therefore converges absolutely if and only if $\Re\delta<\half$, i.e.\ on
\begin{equation}
\Re s<0\,,\quad\Re s'<0\,,\quad 0<\Re\delta<\half\,, \label{sm:eq5}
\end{equation}
which is non-empty for $\Re\nu\ge0$. For $\half\le\Re\delta<1$ the iterated integral still converges at $t\ne t'$, owing to the phase $(t/t')^{\ii w}$, which reflects the kink of the kernel at $t=t'$.

\textit{Proof of Eq.~(2) by closure.}---\textit{Step~1, closing the $s$ contour.} Place the $s'$ line at $\sigma'<\min(0,-\Re\nu)$ and the $s$ line at $\sigma<0$ inside Eq.~(\ref{sm:eq5}). Let $t\le t'$. At fixed $s'$ the $s$ integral converges absolutely and closes to the right. On arcs at $\Re s\simeq M$ the integrand is bounded by $(yt/2)^{2M}/\Gamma(M)^2$. It collects the poles $s=m$ of $\Gamma(-s)$ and $s=s_n\equiv n-\nu-s'$ of $\Gamma(\delta)$, each with residue factor $(-1)^m/m!$ or $(-1)^n/n!$ after the clockwise orientation.

\textit{Step~2, the poles of $\Gamma(-s)$.} At $s=m$ one has $\delta_m=-\nu-m-s'$ and $\Gamma(\delta_m)\Gamma(1-\delta_m)$ $=(-1)^m\Gamma(-\nu-s')\Gamma(1+\nu+s')$, so the residues sum to
\begin{equation}
\sum_{m\ge0}\frac{(yt/2)^{\nu+2m}\,\Gamma(-s')\Gamma(-\nu-s')}{m!\,\Gamma(\nu+m+1)\,(yt'/2)^{-\nu-2s'}}\,,
\end{equation}
i.e.\ $I_\nu(yt)\,\Gamma(-s')\Gamma(-\nu-s')(yt'/2)^{\nu+2s'}$.
With the prefactor $\frac12$ its $s'$ integral is the Barnes integral (\ref{sm:eq1}), giving $I_\nu(yt)K_\nu(yt')$.

\textit{Step~3, the poles of $\Gamma(\delta)$.} At $s=s_n$ one has $\Gamma(1-\delta)=n!$, $\Gamma(-s_n)=\Gamma(\nu+s'-n)$, and $\Gamma(\nu+s_n+1)=\Gamma(n+1-s')$, so the residue is the rational function
\begin{equation}
U_n(s')=\frac{(-1)^n\,(y/2)^{2n}\,t^{\,2n-\nu-2s'}\,t'^{\,\nu+2s'}}{\prod_{j=0}^{n}\bigl[(\nu+s'-n+j)(j-s')\bigr]}\,, \label{sm:eq6}
\end{equation}
whose poles at $s'=j$ and $s'=j-\nu$ ($0\le j\le n$) lie to the right of the $s'$ line. Its remaining $s'$ dependence is $(t'/t)^{2s'}$, which decays as $\Re s'\to-\infty$ for $t<t'$. The $s'$ contour therefore closes to the left on no pole, and $\int\frac{\dd s'}{2\pi\ii}U_n=0$. At $t=t'$ the same follows from $|U_n|\sim|s'|^{-2n-2}$. The sum over $n$ commutes with the $s'$ integral by dominated convergence, since on the line the denominator of $U_n$ exceeds $(n!)^2$ times a quadratic function of $|\Im s'|$.

\textit{Step~4, both orderings.} Hence $\mathcal R=I_\nu(yt)K_\nu(yt')$ for $t\le t'$, and the symmetry of $\mathcal R$ under $(s,t)\leftrightarrow(s',t')$ gives the other ordering, so $\mathcal R=K_\nu(yt_>)I_\nu(yt_<)$. On Eq.~(\ref{sm:eq5}) the order of integration is immaterial, so Eq.~(2) holds there for all $t,t'$. On the rest of the window (\ref{sm:eq3}) it holds for $t\neq t'$, as the iterated integral with the $s$ integral inside, by moving the $s'$ line. The oscillation $(t/t')^{\ii w}$ controls the tails. The ordering is thus carried entirely by the poles of $\Gamma(\delta)$, which contributes nothing on one side of $t=t'$. The reflection formula was used only at the non-integer points $\delta_m$, so integer $\nu$ is included. $\square$

\textit{Light fields.}---For $0<\nu<\thalf$ no further condition is needed. The poles of $\Gamma(-\nu-s')$, kept to the right in Step~1, appear only after the $s$ residues and are not poles of $f$. Equation~(\ref{sm:eq3}) is therefore the window for light fields as well.

\textit{The window of a tree.}---Let $s_{v,l}$ denote the Mellin variable of line $l$ at its end $v$ ($s_l$ or $s_l'$ in Eq.~(3)), so that $\Sigma_v=\sum_{l\ni v}(\nu_l+2s_{v,l})$. The integral (3) is taken on
\begin{equation}
\Re s_{v,l}<0\,,\qquad 0<\Re\delta_l<1\,,\qquad \Re(r_v+\Sigma_v)>0\,. \label{sm:eq7}
\end{equation}
(i)~\textit{Absolute convergence on all of Eq.~(\ref{sm:eq7}).} With $s_{v,l}=\sigma_{v,l}+\ii\tau_{v,l}$, each line factor is bounded as above by powers times $e^{-\pi|\tau_{v_l,l}+\tau_{v_l',l}|}$. Since $|x_v^{-(r_v+\Sigma_v)}|$ is constant, each vertex factor is bounded by powers times $e^{-\frac\pi2|\Im\Sigma_v|}$ with $\Im\Sigma_v=\sum_{l\ni v}(\Im\nu_l+2\tau_{v,l})$. The integrand is therefore bounded by a polynomial times $e^{-\pi\Lambda(\tau)}$, with
\begin{equation}
\Lambda(\tau)=\sum_l\bigl|\tau_{v_l,l}+\tau_{v_l',l}\bigr|+\sum_v\Bigl|\sum_{l\ni v}\tau_{v,l}\Bigr|\,,
\end{equation}
up to bounded shifts. $\Lambda$ is convex and homogeneous of degree one. On a tree it vanishes only at $\tau=0$. A leaf $v$ has a single end, so $\Lambda=0$ forces $\tau_{v,l}=0$ and then $\tau=0$ at the other end of $l$. Removing the leaf repeats the argument. Hence $\Lambda\ge\gamma|\tau|$ with $\gamma>0$, and the integrand decays exponentially in every direction. The anti-diagonal restriction $\Re\delta_l<\half$ of Eq.~(\ref{sm:eq5}) is therefore lifted, and the per-line window widens to $0<\Re\delta_l<1$. (On a cycle, alternating $\tau$ around it gives $\Lambda=0$, the direction of End Matter C along which loop integrands decay only as a power.)

(ii)~\textit{Exchange of time and Mellin integrals.} On the part of Eq.~(\ref{sm:eq7}) with every $\Re\delta_l<\half$ each kernel integral converges absolutely and $\int_0^\infty\dd t\,|t^{r_v+\Sigma_v-1}e^{-x_vt}|=\Gamma(\Re(r_v+\Sigma_v))\,x_v^{-\Re(r_v+\Sigma_v)}$ is constant on the contours, so Fubini's theorem gives Eq.~(3) from Eqs.~(1) and~(2) and the vertex integral. This part is non-empty whenever the window is, since lowering the $a_{v,l}$ of part~(iii) on a line until $\Re\delta_l<\half$ only relaxes the vertex conditions. The rest of the window follows by analyticity in the contour positions.

(iii)~\textit{Straight contours.} With $a_{v,l}=-2\Re s_{v,l}$, Eq.~(\ref{sm:eq7}) reads
\begin{gather}
a_{v,l}>0\,,\qquad 2\Re\nu_l<a_{v_l,l}+a_{v_l',l}<2\Re\nu_l+2\,, \nonumber\\
\sum_{l\ni v}a_{v,l}<r_v+\sum_{l\ni v}\Re\nu_l\,. \label{sm:eq8}
\end{gather}
If every $r_v>0$ and $\Re\nu_l\ge0$, the choice $a_{v,l}=\Re\nu_l+\epsilon$ with $0<\epsilon<\min\bigl(\half,\min_v r_v/\deg(v)\bigr)$ satisfies all three, i.e.\ the straight contours $\Re s_{v,l}=-\half(\Re\nu_l+\epsilon)$, $\Re\delta_l=\epsilon$. On the principal series they involve no mass, and by part~(i) they serve at every kinematic point. Conversely, summing the vertex inequalities over any vertex set $S$ and bounding the left side below by the line inequalities gives the necessary condition $\sum_{v\in S}r_v+\sum_{l\in\partial S}\Re\nu_l>0$, with $\partial S$ the lines with one end in $S$. For $S=\{v\}$ on the principal series this is $r_v>0$. Vertices with $r_v\le0$, whose time integrals diverge at late times, are defined by continuation in $r_v$ on deformed contours.

\textit{The kernel as a Green's function.}---The kernel $\G_\nu$ is the Green's function of the line's mode operator that is regular at early times and, with the factor $(tt')^{3/2}$, vanishes at late times. It differs from the Dirichlet propagator of Ref.~\cite{ArkaniHamed2017} by a boundary term that cancels in every correlator. Hence $\psi_n$ built on $\G_\nu$ is the object from which the observables are assembled. A cut line carries two amputated legs, a mode function and its continued conjugate, with no power of $\eta_0$ (Sec.~\ref{sm:cutting}).

\section{Vertices without external legs}\label{sm:vertex}

At $x_v=0$ the vertex integral $\int_0^\infty\dd t\,t^{r_v+\Sigma_v-1}e^{-x_vt}$ does not exist. Formally, on $\Re(r_v+\Sigma_v)=0$ the substitution $t=e^u$ gives $\int_0^\infty\dd t\,t^{\,r_v+\Sigma_v-1}=2\pi\,\delta\bigl(\Im(r_v+\Sigma_v)\bigr)$. Since $\Sigma_v$ contains $2s_{v,l_0}$ for any line $l_0$ at $v$, the measure $\dd s_{v,l_0}/2\pi\ii=\dd\tau_{v,l_0}/2\pi$ converts it into the weight $\half$ at $r_v+\Sigma_v=0$. The precise statement is the limit $x_v\to0$ of Eq.~(3),
\begin{gather}
\int\!\frac{\dd s_{v,l_0}}{2\pi\ii}\,\frac{\Gamma(r_v+\Sigma_v)\,(\cdots)}{x_v^{\,r_v+\Sigma_v}}\;\xrightarrow{\;x_v\to0\;}\;\half\,(\cdots)\Big|_{s_{v,l_0}=s_*}\,, \label{sm:eq9}\\
s_*=-\half\Bigl(r_v+\sum_{l\ni v}\nu_l\Bigr)-\!\!\sum_{l\ni v,\,l\ne l_0}\!\! s_{v,l}\,, \nonumber
\end{gather}
where the ellipsis is the rest of the integrand.

\textit{Proof.}---In $s_{v,l_0}$ the poles of $\Gamma(r_v+\Sigma_v)$ lie at $s_{v,l_0}=s_*-j/2$, $j\ge0$, to the left of the contour. Move the contour leftward across $s_*$ alone, to $-1<\Re(r_v+\Sigma_v)<0$. Near $s_*$, $\Gamma(r_v+\Sigma_v)\simeq[2(s_{v,l_0}-s_*)]^{-1}$ and $x_v^0=1$, so the crossed residue is $\half(\cdots)|_{s_*}$. On the moved contour $|x_v^{-(r_v+\Sigma_v)}|=x_v^{-\Re(r_v+\Sigma_v)}\to0$, with the integrand dominated by the exponential decay of Sec.~\ref{sm:kernel}, so that integral vanishes. The step is legitimate when the moved contour and the residue point stay inside the rest of Eq.~(\ref{sm:eq7}). This holds when straight contours obeying the other conditions reach the plane $\Re(r_v+\Sigma_v)=0$. $\square$

\textit{Range of $r_v$.}---On the principal series $\Re\Sigma_v=-\sum_{l\ni v}a_{v,l}$. We work on the kernel window $0<\Re\delta_l<\half$ of Eq.~(\ref{sm:eq5}), where Eq.~(3) follows from Fubini's theorem [part~(ii) of Sec.~\ref{sm:kernel}]. There $0<a_{v,l}<1$, so $\Re\Sigma_v$ ranges over $(-\deg(v),0)$ when the far-end variables are free to approach zero. This holds when every neighbor of $v$ carries external legs. The plane is then reached for
\begin{equation}
0<r_v<\deg(v)\,, \label{sm:eq10}
\end{equation}
which therefore suffices. A vertex without external legs joining three, four, or five scalar lines has $r_v=\thalf\deg(v)-3=\thalf,3,\frac92$ and satisfies Eq.~(\ref{sm:eq10}). In the correlator the rule applies block by block, with amputated legs represented by Eq.~(\ref{sm:eq1}), whose $a_{v,l}$ is unbounded above.

\textit{Triple exchange.}---Let the central vertex carry the ends $s_l'$ of the lines $l\in\{A,B,C\}$, whose outer ends sit at mixing vertices with $r=\half$ and energies $x_l$. With the leg function
\begin{align}
M_l(s')&=\frac{\Gamma(-s')}{\Gamma(\nu_l+s'+1)}\int\!\frac{\dd s}{2\pi\ii}\,\frac{\Gamma(-s)\Gamma(\delta_l)\Gamma(1-\delta_l)}{\Gamma(\nu_l+s+1)} \nonumber\\
&\quad\times\Bigl(\frac{y_l}2\Bigr)^{-2\delta_l}\Gamma\bigl(\tfrac12+\nu_l+2s\bigr)\,x_l^{-(\frac12+\nu_l+2s)}\,,
\end{align}
$\delta_l=-\nu_l-s-s'$, the rule (\ref{sm:eq9}) solved for $s_C'$ gives
\begin{gather}
\psi_{\rm triple}=\frac1{16}\iint\!\frac{\dd s_A'\,\dd s_B'}{(2\pi\ii)^2}\,M_A(s_A')M_B(s_B')M_C(s_C')\,, \label{sm:eq11}\\
s_C'=-\tfrac34-\tfrac12(\nu_A+\nu_B+\nu_C)-s_A'-s_B'\,, \nonumber
\end{gather}
(the $\frac1{16}$ is $(\half)^3$ from the kernels times $\half$ from the vertex).

\textit{The mass insertion.}---A two-line vertex without legs, $\deg(v)=2$ and $r_v=0$, sits at the edge of Eq.~(\ref{sm:eq10}). Reaching $\Re\Sigma_v=0$ then requires $a_{v,l}\to0$ on both lines, which pinches the two contours against the poles of $\Gamma(-s_{v,l})$ at the origin. The displacement crosses one double pole, where the constraint pole meets the first pole of a $\Gamma(-s)$. In the time domain the vertex integrand behaves as $t^{2\nu-1}$ at small $t$, and the integral is its Hadamard finite part. Differentiating the Green's equation $(\theta^2-\nu^2-y^2t^2)\G_\nu=-t'\delta(t-t')$, $\theta=t\,\dd/\dd t$, in $m^2$ with $\nu^2=\frac94-m^2$ gives, for $\Re\nu>0$, and on the principal series as its continuation,
\begin{equation}
\frac{\partial\G_\nu(y;t,t')}{\partial m^2}=\int_0^\infty\frac{\dd t''}{t''}\,\G_\nu(y;t,t'')\,\G_\nu(y;t'',t')\,,
\end{equation}
and $\dd t''/t''$ is the vertex with $r_v=0$, so the finite part equals the mass derivative of the single-line diagram. In Mellin--Barnes form the result is the rule (\ref{sm:eq9}) plus the residue of the crossed double pole, a boundary term that factorizes between the two outer vertices. On the principal series it is always present, and only the sum is symmetric under exchange of the outer vertices.

\section{Feynman kernel and the cutting rule}\label{sm:cutting}

\textit{Modes and propagators.}---The Bunch--Davies mode of a line of energy $y$ is $u=N_\nu(-\eta)^{3/2}H^{(1)}_\nu(-y\eta)$ with $N_\nu=\frac{\sqrt\pi}2e^{\ii\pi(2\nu+1)/4}$, i.e.\ $N_{\ii\mu}=\frac{\sqrt\pi}2e^{-\pi\mu/2}e^{\ii\pi/4}$, and its conjugate continues to $u^*\propto H^{(2)}$. A vertex on branch $a=\pm$ is rotated to $\eta=a\,\ii t$, $t>0$, as the prescriptions $\eta\to-\infty(1\mp\ii\epsilon)$ allow. On the forward branch the modes are
\begin{align}
u_\nu^*(\ii t)&=-\frac{\ii}{\sqrt\pi}\,t^{3/2}K_\nu(yt)\,, \nonumber\\
u_\nu(\ii t)&=-\ii\sqrt\pi\,t^{3/2}\Bigl[I_\nu-\frac\ii\pi e^{\ii\pi\nu}K_\nu\Bigr](yt)\,, \label{sm:eq12}
\end{align}
consistent with the canonical Wronskian $u\,\partial_\eta u^*-u^*\partial_\eta u=\ii\eta^2$ through $z(I_\nu K_\nu'-I_\nu'K_\nu)=-1$. Schwarz reflection gives $u(-\ii t)=\overline{u^*(\ii t)}$, $u^*(-\ii t)=\overline{u(\ii t)}$. With $u$ at the later time (smaller $t$) the four propagators are
\begin{align}
G_{++}(t,t')&=u(\ii t_<)\,u^*(\ii t_>)=-(tt')^{3/2}\,G^F_\nu(y;t,t')\,, \label{sm:eq13}\\
G_{+-}(t,t')&=u^*(\ii t)\,\overline{u^*(\ii t')} \nonumber\\
&=\frac1\pi(tt')^{3/2}K_\nu(yt)K_\nu(yt')\,, \label{sm:eq14}
\end{align}
with $G_{--}=\overline{G_{++}}$, $G_{-+}=G_{+-}$, and $t$ the forward time in $G_{+-}$. Equation~(\ref{sm:eq13}) with $e^{\ii\pi\nu}=e^{-\pi\mu}$ at $\nu=\ii\mu$ gives the Feynman kernel of the main text,
\begin{equation}
\begin{split}
G^F_\nu(y;t,t')&=K_\nu(yt_>)\Bigl[I_\nu-\frac\ii\pi e^{-\pi\mu}K_\nu\Bigr](yt_<)\\
&=\frac{e^{\pi\mu}\G_\nu-e^{-\pi\mu}\G_{-\nu}}{2\sinh\pi\mu}\,,
\end{split} \label{sm:eq15}
\end{equation}
where the second form follows from $K_{-\nu}=K_\nu$ and $I_{-\nu}=I_\nu+\frac2\pi\sin(\pi\nu)K_\nu$ with $\sin(\ii\pi\mu)=\ii\sinh\pi\mu$. Since $\Re(\pm\ii\mu)=0$, both terms are represented by Eq.~(2) on the same contours, and the amputated leg $K_\nu$ of Eq.~(\ref{sm:eq14}) by Eq.~(\ref{sm:eq1}). Every block below is therefore a Mellin--Barnes integral of the form of Eq.~(3). The cut-line factor (\ref{sm:eq14}) carries $(-\ii/\sqrt\pi)(\ii/\sqrt\pi)=1/\pi$ from its two amputated ends and no power of $\eta_0$.

\textit{Vertices.}---A vertex on branch $a$ collects $\ii a\,g_v$ and the measure $\sqrt{-g}\,\dd\eta=\eta^{-4}\dd\eta$, which becomes $-\ii a\,t^{-4}\dd t$ on $t\in(0,\infty)$. Per conformally coupled leg it also collects the bulk-to-boundary factor, $u_k(\eta_0)u_k^*(\eta)$ on the forward branch and its conjugate on the backward one, which tends to $(\eta_0/2k)\,(\ii a)\,t\,e^{-kt}$. With $g_v$ and $\eta_0/2k$ per leg stripped it carries $c_v(a)\,t^{n_v-4}e^{-x_vt}$ with $c_v(a)=(\ii a)^{n_v}$. Massless or $\pid$ legs change only the power of $t$ and the phase, and the proof uses only
\begin{equation}
c_v(-a)=\overline{c_v(a)}\,. \label{sm:eq16}
\end{equation}
The stripped correlator is the branch sum
\begin{align}
\mathcal A=\sum_{a\in\{\pm\}^n}\int_0^\infty&\prod_v\dd t_v\,c_v(a_v)\,t_v^{n_v-4}e^{-x_vt_v} \nonumber\\
&\times\prod_{l}G_{a_{v_l}a_{v_l'}}(t_{v_l},t_{v_l'})\,. \label{sm:eq17}
\end{align}

\textit{Theorem.}---Call a line cut if its ends lie on different branches. Then
\begin{equation}
\mathcal A=\sum_{\mathcal C}\;\sum_{\varepsilon}\;\prod_B\mathcal B_B^{(\varepsilon_B)}\,, \label{sm:eq18}
\end{equation}
with $\mathcal B_B^{(+)}=\mathcal B_B$ and $\mathcal B_B^{(-)}=\mathcal B_B^*$.
Here $\mathcal C$ runs over sets of internal lines, and $B$ over the blocks (connected components) left when the lines of $\mathcal C$ are severed. The signs $\varepsilon_B=\pm$ run over the allowed choices of branch for each block, forward or backward, in which blocks joined by a cut line lie on opposite branches, and
\begin{align}
\mathcal B_B&=\Bigl(\prod_{v\in B}c_v(+)\Bigr)(-1)^{L_B}\Bigl(\frac{-\ii}{\sqrt\pi}\Bigr)^{c_B}\psi_B^F\,, \label{sm:eq19}\\
\psi_B^F&=\int_0^\infty\prod_{v\in B}\dd t_v\,t_v^{\,r_v-1}e^{-x_vt_v} \nonumber\\
&\quad\times\prod_{\substack{l\in B\\\text{uncut}}}G^F_{\nu_l}(y_l;t_{v_l},t_{v_l'})\prod_{\substack{\text{cut ends}\\(l,v),\,v\in B}}K_{\nu_l}(y_lt_v)\,, \nonumber
\end{align}
with $L_B$ uncut lines and $c_B$ cut ends in $B$.

\textit{Proof.}---\textit{Step~1, bijection.} An assignment $a$ fixes the cut set $\mathcal C(a)=\{l:a_{v_l}\neq a_{v_l'}\}$. Uncut lines join vertices on one branch, so each block of $\mathcal C(a)$ lies on a single branch $\varepsilon_B(a)$, and blocks joined by a cut line lie on opposite branches. Conversely, a cut set with an allowed choice of branch for each block defines an assignment. The map $a\mapsto(\mathcal C(a),\varepsilon(a))$ is thus a bijection onto cut sets with allowed branch choices, and the sum (\ref{sm:eq17}) becomes the double sum (\ref{sm:eq18}).

\textit{Step~2, factorization.} By Eq.~(\ref{sm:eq14}) a cut line is a product $u^*(\ii t_v)\,\overline{u^*(\ii t_{v'})}$ of a function of each of its times, with $v$ on the forward side. At fixed $a$ the integrand of Eq.~(\ref{sm:eq17}) therefore factorizes over blocks, a forward block carrying $c_v(+)$, $G_{++}$ on its uncut lines, and one amputated leg $u^*(\ii t_v)$ per cut end.

\textit{Step~3, backward blocks.} A backward block carries $c_v(-)$, $G_{--}$, and $\overline{u^*(\ii t_v)}$, on a real measure with real exponentials, so by Eqs.~(\ref{sm:eq13}) and (\ref{sm:eq16}) it is the complex conjugate of the same block on the forward branch.

\textit{Step~4, normalization.} In a forward block each uncut line gives $-(tt')^{3/2}G^F_\nu$ by Eq.~(\ref{sm:eq13}), and each cut end $-\frac{\ii}{\sqrt\pi}t^{3/2}K_\nu$ by Eq.~(\ref{sm:eq12}). With every line end supplying $t_v^{3/2}$, cut ends included, $t_v^{n_v-4}t_v^{\frac32\deg(v)}=t_v^{\,r_v-1}$. This is Eq.~(\ref{sm:eq19}). $\square$

The proof never uses that the diagram is a tree, so Eq.~(\ref{sm:eq18}) holds at fixed loop momenta. A cut set that severs a cycle only once contributes nothing, since both ends of the severed line then lie in one block. The rotation of each branch assumes that every block converges at small $t_v$, where $G^F_\nu$ and an amputated leg both behave as $t_v^{-\Re\nu_l}$. This means $\sum_{l\ni v}\Re\nu_l<r_v$ at every vertex, which on the principal series is $r_v>0$.

\textit{The cutting rule for trees.}---For a tree every line separates the diagram, so the blocks of any cut set are linked by the cut lines without forming a loop. Choosing the branch of one block then fixes all the others, which leaves exactly two allowed choices, conjugate to each other. Their sum is $2\Re$ of either term. Every cut line joins a forward and a backward block and contributes $(-\ii/\sqrt\pi)(\ii/\sqrt\pi)=1/\pi$, and the line signs multiply to $\prod_B(-1)^{L_B}=(-1)^{L-n_c}$ for $n_c$ cut lines. When the vertex phases $c_v(+)$ are real, as for the $\pid$ couplings of the main text ($c_v(+)=1$), this gives
\begin{align}
\mathcal A&=(-1)^L\,2\Re\psi_n^F \nonumber\\
&\quad+\sum_{\mathcal C\neq\emptyset}(-1)^{L-n_c}\,\frac{2}{\pi^{n_c}}\,\Re\Bigl[\prod_B\bigl(\psi_B^F\bigr)^{(\varepsilon_B)}\Bigr]\,, \label{sm:eq20}
\end{align}
the rule stated in the main text, with the Feynman kernel on every uncut line and the blocks on the two sides of each cut line conjugated relative to each other. With $L=1$ it reduces to the bracket of Eq.~(\ref{sm:single}), $-2\Re\psi_2^F+\frac2\pi\Re(\Phi_A\Phi_B^*)$. For complex vertex phases, e.g.\ $c_v(+)=\ii^{n_v}$ with odd $n_v$, as for the triangle with one conformally coupled leg at each vertex, each block keeps its own phase inside Eq.~(\ref{sm:eq18}), conjugated with the block. When every $n_v$ is even these phases are real and multiply to $\ii^N$.

\section{Single and double exchange}\label{sm:double}

This section collects the conventions of the single exchange, the squeezed limit of the trees with the moment form of $\cM_3$ (whose closed form is Eq.~(11) of End Matter~B), and the double exchange. Every coupling enters $\Lint$ with a plus sign.

\textit{Single exchange.}---Single exchange with two conformally coupled legs at each vertex is the single-line case of Eq.~(3) at $r_v=\half$~\cite{Sleight2020,SleightTaronna2020}. The two branches of the discontinuity across its partial-energy cut carry the relative weight $e^{-2\pi\mu}$, a thermal ratio at the Gibbons--Hawking temperature $T=1/2\pi$~\cite{ArkaniHamed2015}. Its square root $e^{-\pi\mu}$, an interference amplitude rather than a pair-production probability, sets the collider signal~\cite{ArkaniHamed2015,Lee2016,Tong2022,QinXianyu2023loop}. This counting returns in the squeezed limits.

Consider the single-exchange bispectrum of the inflaton fluctuation $\pi$ coupled to a heavy field $\sigma$ through $\Lint=a^3(\rho\,\pid\sigma+\kappa\,\pid^2\sigma)$. All results are at leading order in the couplings, in particular in the mixing, $\rho\ll1$. At $\rho\gtrsim1$ the mixing must be resummed~\cite{Kumar2026scalars,Philcox2026lim,Huenupi2026linear} and the squeezed exponent is set by an effective mass~\cite{Wang2026strong,Huenupi2026,Pinol2026,Belrhali2026strong}. The power spectrum of $\pi$ is $P(k)=1/(2k^3)$, and each $\pid$ leg brings one factor $k_a^2$. A prime strips the momentum-conserving delta function, and $k_{12}=k_1+k_2$. The assembly gives
\begin{align}
\langle\pi_{\bk_1}\pi_{\bk_2}\pi_{\bk_3}\rangle' &= 2\rho\kappa\prod_a k_a^2P(k_a)\sum_{\rm cyc}\Bigl[-2\Re\psi_2^F(k_{12},k_3;k_3) \nonumber\\
&\qquad + \tfrac{2}{\pi}\Re\bigl(\Phi_A\,\Phi_B^*\bigr)\Bigr]\,, \label{sm:single}
\end{align}
where the sum runs over the three assignments of $\bk_3$ to the $\rho$ vertex. The blocks $\Phi_A$ and $\Phi_B$ are the two halves of the cut diagram, $\Phi_A$ at $x=k_{12}$ and $\Phi_B$ at $x=y=k_3$. 

The factor 2 multiplying the cut term in Eq.~(\ref{sm:single}) counts the two conjugate branch assignments. The stripped phase $\ii^N$ of $\psi^{\rm full}$ multiplies each term inside the real part when every vertex phase is real, as for $\pid$ legs. With an odd number of conformally coupled legs at a vertex, as in the triangle with $r_v=1$, the vertex phases are complex, and each block keeps its own phase under the conjugation (Sec.~\ref{sm:cutting}). In Eq.~(\ref{sm:single}), $\psi_2^F$ is the single-line case of Eq.~(3) with the Feynman kernel at $r_A=\frac52$ for the $\kappa$ vertex and $r_B=\half$ for the $\rho$ vertex, since each $\pid$ leg contributes $t^2e^{-kt}$. The blocks $\Phi_A$, $\Phi_B$ carry the same powers with the severed end's $t^{3/2}$ counted inside, and the 2 in $2\rho\kappa$ counts the exchange of the two $\pid$ legs. More generally, the bispectrum of one leg assignment of any tree here is the product of the couplings, $\prod_ak_a^2P(k_a)$, and $\mathcal A$ of Eq.~(\ref{sm:eq20}), with a factor 2 for each pair of legs sharing a vertex.

\textit{Squeezed limit.}---Consider the bispectra at $k_1=k_2$ with the mixing leg $\bk_3$ squeezed, $k_3/k_1\to0$. The leading residue of the soft line's $\Gamma(-s')$ in Eq.~(3) and Barnes' first lemma at the soft mixing vertex give
\begin{equation}
S = \San + 2\Re\Bigl[\cM_L\Bigl(\frac{k_3}{k_1}\Bigr)^{\frac12+\ii\mu}\Bigr]\Bigl(1+O\Bigl(\frac{k_3^2}{k_1^2}\Bigr)\Bigr), \label{sm:squeezed}
\end{equation}
with $L=2,3$ the number of lines, $\San$ analytic in $k_3/k_1$, and $\mu$ the soft line's mass. Define the thermal cut weights $\lambda_\pm(\mu)=\pm(e^{\pm\pi\mu}\mp\ii)/(2\sinh\pi\mu)$ and $c(\mu)=1-\ii/\sinh\pi\mu$, the unit-energy mixing block $\hat\Phi_l=\sqrt{\pi/2}\,\Gamma(\half{+}\ii\mu_l)\,\Gamma(\half{-}\ii\mu_l)$, and the hard line dressed through the Feynman kernel, $\F_l(t)=\int_0^\infty\dd u\,u^{-1/2}e^{-u}\,G^F_{\nu_l}(1;u,t)$. The triple exchange has, in moment form,
\begin{align}
\cM_3 &= -\frac{2^{-\ii\mu_C}\,\hat\Phi_C}{8\,\Gamma(1{+}\ii\mu_C)}\Bigl[\lambda_+\hat\cT_+ + \lambda_-\overline{\hat\cT_-} \nonumber\\
&\qquad + \tfrac{1}{\pi^2}\,c\,\hat\Phi_A\hat\Phi_B\,\hat T_{KK}(\ii\mu_C)\Bigr], \label{sm:M3mom}
\end{align}
with $\lambda_\pm$ and $c$ at $\mu_C$, overlines denoting complex conjugation, and $\hat T_{KK}(p)=2^{p-3/2}\,\Gamma(\thalf{+}p)^{-1}\prod_{\pm\pm}\Gamma\bigl(\half(\thalf+p\pm\ii\mu_A\pm\ii\mu_B)\bigr)$. The remaining coefficients are $\hat\cT_\pm=\bigl[\hat T_{FF}-\frac1\pi\hat\Phi_A\hat U-\frac1\pi\hat\Phi_B\hat V\bigr](\pm\ii\mu_C)$, where $\hat T_{FF}(p)=\int_0^\infty\dd t\,t^{1/2+p}\,\F_A(t)\F_B(t)$, $\hat U(p)=\int_0^\infty\dd t\,t^{1/2+p}K_{\ii\mu_A}(t)\,\F_B(t)$, and $\hat V=\hat U$ with $A\leftrightarrow B$ are the central-vertex moments. At general masses they close as single sums of products of ${}_3F_2$ at unit argument, Eqs.~(10) and~(11) and Sec.~\ref{sm:closed}. The double-exchange coefficient $\cM_2$ is given below.

\textit{The triangle at leading order.}---In the triangle the hard factor of $\cM_{\rm tri}=\cM^{(0)}_+$ is the integral of $t^p\,t'^{\,p'}e^{-t-t'}\,\G_{\nu_A}(1;t,t')$, with the powers $p$, $p'$ deposited by the soft lines. This is the doubly folded limit of the single-exchange seed integral of Ref.~\cite{QinXianyu2023closed} [its Eqs.~(92)--(93)]. Hence at general masses $\cM_{\rm tri}$ is given by Gamma functions and one ${}_3F_2$ at unit argument. When the soft line at a vertex is degenerate with the hard line, $p=1+\nu_A$, the series reduces to $(p'+\nu_A+3)\,\Gamma(p'+\nu_A+1)/[2^{p'+\nu_A+2}(2\nu_A+3)]$. Then $\cM_{\rm tri}$ is a ratio of Gamma functions for three equal masses, where it coincides with Eq.~(26) of Ref.~\cite{QinXianyu2026triangle} to $10^{-16}$, and also for $\mu_A=\mu_B\ne\mu_C$.

\begin{figure}[t]
\centering
\includegraphics[width=0.6\textwidth]{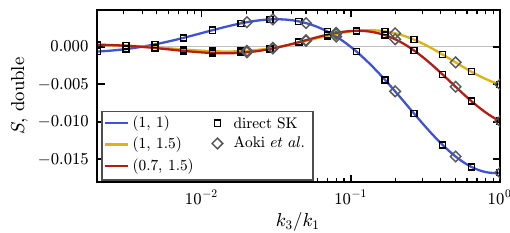}
\caption{Double-exchange bispectrum $S=(k_1k_2k_3)^2\,\langle\pi\pi\pi\rangle'$ for the channel with $\bk_1$ on the cubic vertex $\pid\sigma_1\sigma_2$ and $\sigma_1$, $\sigma_2$ mixing into $\bk_2$, $\bk_3$, at unit couplings and $k_1=k_2$. Labels give $(\mu_1,\mu_2)$, with the line on the $\bk_3$ vertex last. Lines are Eq.~(3), squares direct Schwinger--Keldysh integration, and diamonds the series of Ref.~\cite{Aoki2024} truncated at $N=80$ terms.}
\label{fig:S1}
\end{figure}

\textit{The double exchange.}---In the double exchange two principal-series fields $\sigma_1$, $\sigma_2$ couple to the inflaton through mixing vertices $\rho_l\,\pid\sigma_l$ ($r_v=\half$) and are joined by the cubic vertex $\pid\sigma_1\sigma_2$ ($r_v=2$). The diagram is evaluated from Eq.~(3) and assembled by the cutting rule applied to both lines (Sec.~\ref{sm:cutting}). It is known from Ref.~\cite{Aoki2024} and serves as the simplest check of the tree construction. Figure~\ref{fig:S1} shows it on isosceles triangles.

\textit{Comparison with Ref.~\cite{Aoki2024}.}---The series solution of the bootstrap equation in Ref.~\cite{Aoki2024} [its Eqs.~(4.2), (4.4)--(4.6), and (4.8)] is a single sum, which we order in powers of the soft-leg ratio $2k_3/(k_1{+}k_2{+}k_3)$. We truncate it at $N=80$ terms, where the omitted remainder is below $10^{-13}$ relative at every point. It then reproduces Fig.~\ref{fig:S1} at the 21 points shown, $k_3/k_1\ge0.02$, with maximum and median differences $3\times10^{-11}$ and $5\times10^{-12}$. At half the default grid step of the line blocks they fall to $2\times10^{-12}$ and $3\times10^{-13}$. The number of terms needed for $10^{-13}$ grows from 8 at $k_3/k_1=0.02$ to 74 at the equilateral point. Ordered in the hard-leg ratio $2k_1/(k_1{+}k_2{+}k_3)$ instead, the same series needs more than 3000 terms at $k_3/k_1=0.02$ (the slow convergence as this ratio approaches 1 that restricts the shape comparison of Ref.~\cite{Aoki2024} to $k_3/k_1\gtrsim0.15$ along this line). Direct Schwinger--Keldysh integration covers the full range, down to $k_3/k_1=2\times10^{-3}$.

\textit{Squeezed coefficient.}---With $\lambda_\pm$, $c$, $\hat\Phi_l$, and $\F_l$ as defined above, the double exchange has, in Eq.~(\ref{sm:squeezed}),
\begin{align}
\cM_2 &= \frac{2^{-\ii\mu_2}\,\hat\Phi_2}{8\,\Gamma(1{+}\ii\mu_2)}\Bigl[\lambda_+\hat H_F(2{+}\ii\mu_2)+\lambda_-\overline{\hat H_F(2{-}\ii\mu_2)} \nonumber\\
&\qquad - \tfrac1\pi\,c\,\hat\Phi_1\,\hat H(2{+}\ii\mu_2)\Bigr], \label{sm:M2}
\end{align}
with $\lambda_\pm$ and $c$ at $\mu_2$, $\hat H(a)=\sqrt\pi\,2^{-a}\,\Gamma(a{+}\ii\mu_1)\Gamma(a{-}\ii\mu_1)/\Gamma(a{+}\half)$, and $\hat H_F(a)=\int_0^\infty\dd t\,t^{a-1}e^{-t}\,\F_1(t)$ the same moment of the hard line dressed through the Feynman kernel. At the mixing vertex's kinematics $x=y$ the dressed line reduces not to Gamma functions but to the ${}_3F_2$ series at unit argument of Ref.~\cite{QinXianyu2023closed}. For comparison, one channel of the single exchange of Eq.~(\ref{sm:single}) at $\rho=\kappa=1$ has $\cM_1=-4^{-\ii\mu}\,c(\mu)\,\hat\Phi\,\Gamma(\tfrac52{+}\ii\mu)/[2^{9/2}\,\Gamma(1{+}\ii\mu)]$, with $\hat\Phi$ the mixing block at the soft mass $\mu$. At $\mu=1$ on every line and with all couplings set to one in units of $H$ (the physical ratios carry the couplings), $|\cM_2/\cM_1|=0.371$ and $|\cM_3/\cM_1|=0.317$. To leading order in the Boltzmann factor, the squeezed double-exchange signal at unequal masses was obtained in closed form, with a ${}_3F_2$, in Ref.~\cite{Aoki2024} [its Eqs.~(4.21)--(4.22)]. Equation~(\ref{sm:M2}) keeps all orders in the cut weights. The two agree at leading order. At equal masses Eq.~(\ref{sm:M2}) equals $\frac18\,c(\mu)$ times the coefficient in Eq.~(4.20) of Ref.~\cite{Aoki2024}, up to relative corrections of order $e^{-2\pi\mu}$. The factor $\frac18$ converts normalizations, since Ref.~\cite{Aoki2024} sums the two leg assignments and the two contractions of its vertex $\lambda\,\sigma^2\varphi'$ and writes the signal as $\Re$ rather than $2\Re$. The cut weight $c=1-\ii/\sinh\pi\mu$ is the first correction beyond leading order, of relative size $2e^{-\pi\mu}$. At unequal masses the factor $\frac18$ alone relates Eq.~(\ref{sm:M2}) to Eq.~(4.22) of Ref.~\cite{Aoki2024}, whose ${}_3\mathcal F_2$ is the dressed hypergeometric function. The relative differences are of order $e^{-\pi\min(\mu_1,\mu_2)}$, e.g.\ $2\times10^{-4}$ with soft mass $\mu_2=3$ and hard mass $\mu_1=4$, and $4\times10^{-4}$ with $\mu_2=4$ and $\mu_1=3$. At equal masses $\cM_2$ and $\cM_3$ equal $4^{-\ii\mu}C^{(2)}_{\cS+}/64$ and $4^{-\ii\mu}C^{(3)}_{\cS+}/32$ in the notation of Ref.~\cite{XianyuZang2026} (with $C^{(3)}_{\cS+}$ normalized as in its large-mass limit, Eq.~(111), and Fig.~8), with one leg assignment per channel and $\pi$ the canonically normalized inflaton fluctuation. At fixed soft mass the dependence on the hard lines' masses is Boltzmann-free, falling roughly as $e^{-\mu}$ per hard line over $0.5\le\mu\le2$ and more slowly beyond.

\section{Pole terms of the bubble and the triangle}\label{sm:poles}

\textit{The bubble.}---Inserting Eq.~(5) into Eq.~(3) replaces $\prod_l(y_l/2)^{-2\delta_l}$ by $2^{2\cS}k^{3-2\cS}C_2(\delta_1,\delta_2)$, with $k$ the momentum through the loop, and gives a four-variable Mellin--Barnes integral $\psi_2^{\rm MB}$ on contours with $\half<\Re\cS<\thalf$. Let $h_2(\cS)$ be $\psi_2^{\rm MB}$ with $k^{3-2\cS}C_2$ omitted and the $\cS$ integration left undone. Along the direction of End Matter~C its integrand decays only as $|\Im s|^{2\cS-4}$, and at $\cS=\thalf$ this tail diverges, so $h_2$ has a simple pole there, with finite part FP. The pole comes from the large-momentum behavior $K_\nu I_\nu\to1/2yt$ of both lines, and its residue fixes the coefficient $\pi/(x_1{+}x_2)$ of the logarithm. Cut off at comoving $|\bl|=\Lambda$, the loop integral is
\begin{equation}
\psi_2^{\rm loop}\big|_\Lambda = \psi_2^{\rm MB} + \frac{\pi\ln\Lambda}{x_1+x_2} + 2\pi\FP_{\cS=\frac32}h_2 + o(1)\,, \label{sm:bubble}
\end{equation}
whose third term is independent of $k$. The divergence returns as the residue $2\pi$ of the one pole crossed, with a mass-independent coefficient and the energy dependence of a contact term, removed by one quartic counterterm of the external legs. Closing the contours gives the analytic background together with the nonlocal and local signals, residue series in $(k^2/4x_1x_2)^{\pm(\nu_1+\nu_2)}$ and $(x_2/x_1)^{\pm(\nu_1+\nu_2)}$ respectively, which converge only for $k<x_2<x_1$. The sum of single exchanges of Sec.~\ref{sm:closed} converges at every shape short of the folded limit.

\textit{The conformally coupled triangle.}---With conformally coupled legs ($r_v=1$) and a cutoff $\Lambda$ on $|\bl|$ the loop-integrated coefficient is a finite Mellin--Barnes integral plus a logarithm with a mass-dependent coefficient $D_0$ and a remainder $R$,
\begin{align}
\psi_3^{\rm loop} &= \prod_l\biggl[\frac12\iint\!\frac{\dd s_l\dd s_l'}{(2\pi\ii)^2}\,
\frac{\Gamma(-s_l)\Gamma(-s_l')\Gamma(\delta_l)\Gamma(1-\delta_l)}{\Gamma(\nu_l{+}s_l{+}1)\Gamma(\nu_l{+}s_l'{+}1)}\,4^{\delta_l}\biggr] \nonumber\\
&\quad\times\prod_v\Gamma(1{+}\Sigma_v)\,k_v^{-(1+\Sigma_v)}\,C_3(\delta;k) \nonumber\\
&\quad + 4\pi D_0\ln\Lambda + R\,, \label{sm:triangle}
\end{align}
with total exponent $\cS=\delta_A+\delta_B+\delta_C$. We take the first term on contours with $\half<\Re\cS<\thalf$, and the last two are the residue of the pole of $\Gamma(\cS-\thalf)$ crossed in continuing from the loop integral's domain $\Re\cS>\thalf$. Regulating the measure as $\dd^3\ell\,\ell^{-2\epsilon}$ shifts $\delta_C\to\delta_C+\epsilon$ inside $C_3$ only and opens the window $\thalf-\epsilon<\Re\cS<\thalf$, where loop integral and contours commute. Deforming back to contours with $\Re\cS>\half$ crosses only that pole, whose residue is $\Res_{\cS=3/2}C_3=\pi^{3/2}/\Gamma(\thalf)=2\pi$, independent of masses and momenta. Writing $h(\cS)$ for the first term of Eq.~(\ref{sm:triangle}) with $C_3$ replaced by one and the contour of $\delta_A$ removed by the constraint $\delta_A{+}\delta_B{+}\delta_C=\cS$, the regulated coefficient is the first term plus $2\pi\,h(\thalf-\epsilon)$. Taking the residues of two vertex factors at $r_v+\Sigma_v=0$ leaves the third as $\Gamma(3-2\cS)\,x_v^{2\cS-3}$, since $\sum_v(r_v+\Sigma_v)=3-2\cS$. Hence $h$ has a simple pole there,
\begin{equation}
h(\cS) = -\frac{D_0}{\cS-\frac32}+h_{\rm r}+O\bigl(\cS-\tfrac32\bigr), \qquad R=2\pi h_{\rm r}\,, \label{sm:hpole}
\end{equation}
with $h_{\rm r}$ the finite part and $D_0$ the three-variable Mellin--Barnes integral that remains at $\cS=\thalf$ after all three vertex residues have stripped their powers of $x_v$. This integral is therefore independent of the momenta and equal to the unit-energy triple time integral, Eq.~(14). The integrand's large-$\ell$ tail is $D_0/\ell^3+O(\ell^{-4})$, and for such a tail the analytic regulator $\ell^{-2\epsilon}$ and a sharp cutoff at $\ell=\Lambda$ give identical finite parts. Trading $1/2\epsilon$ for $\ln\Lambda$ therefore gives Eq.~(\ref{sm:triangle}) with no finite mismatch. A regulator scale $\ell_0\neq1$ shifts $R$ by $4\pi D_0\ln\ell_0$ (Sec.~\ref{sm:checks}). The finite part $h_{\rm r}$ retains continuous powers of $x_v/x_{v'}$, so $R$ alone is not a contact term. Only its sum with the first term of Eq.~(\ref{sm:triangle}) takes the form $-4\pi D_0\ln k_T+b$, which conformal symmetry fixes (it is broken only by the scale anomaly of three dimension-two operators)~\cite{Bzowski2016}. The logarithm is secular, not ultraviolet. It grows as $4\pi D_0\ln\bigl(a(\eta_0)H/k_T\bigr)$ with $k_T=k_1+k_2+k_3$, requires no bulk counterterm, and cancels in the equal-time correlator (Sec.~\ref{sm:closed}).

\section{Numerical checks}\label{sm:checks}

The tree-level results, the bubble and triangle evaluations of Figs.~\ref{fig:S3} and~\ref{figS:loops}, and their comparisons use the code of Ref.~\cite{McCuneCode} (release v1.0). The closed forms of Sec.~\ref{sm:closed} and the strong-mixing results of Sec.~\ref{sm:strong} use further scripts by the author, to be added to that code. Relative differences are quoted. Contour integrals are evaluated by trapezoidal quadrature on straight contours inside the stated windows and spot-checked at 25 digits. With this code one general-kinematics triple-exchange bispectrum, for one leg assignment, takes 0.1\,s on one core after about 0.5\,s of preparation per set of masses. This timing is at a faster setting than that of the checks in End Matter D. Its largest error over the validation set of the next paragraph is $7\times10^{-11}$ for $\mu=(1,1.5,0.7)$, $1.3\times10^{-10}$ for equal masses $\mu=1$ and 2, and $5\times10^{-10}$ overall. That set has 198 configurations and 201 diagram points, since for three double-exchange mass pairs the equilateral point enters in both squeezing modes. CosmoFlow~\cite{Werth2024} takes minutes per configuration.

\begin{figure}[t]
\centering
\includegraphics[width=0.6\textwidth]{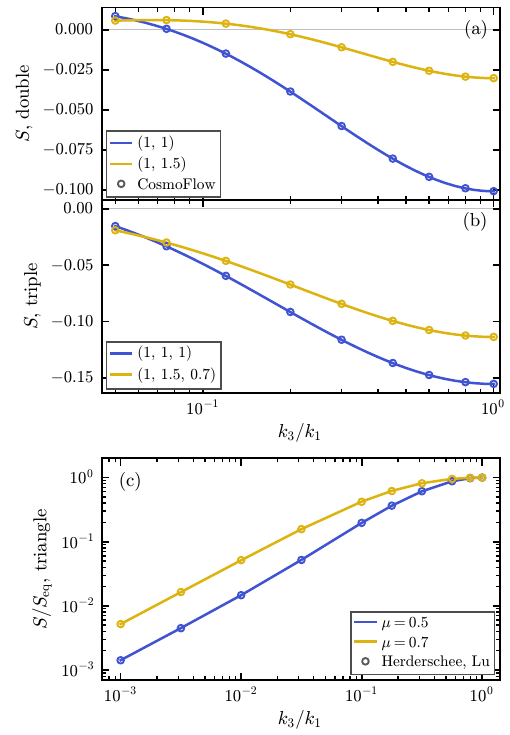}
\caption{Multi-exchange and triangle diagrams against external codes. (a)~Double- and (b)~triple-exchange correlators $S(1,1,k_3)$, summed over the six leg assignments, from Eq.~(3) (lines) and from CosmoFlow~\cite{Werth2024} run unmodified at $\rho_l=0.02$ (circles) at the mass parameters indicated. They agree to a median 0.1\%. (c)~The finite triangle with $\pid$ legs at equal masses $\mu$ along $k_1=k_2$, normalized at the equilateral point. Lines through the computed points are brute-force integration over the vertex times and the loop momentum, the method that checks the representation in Fig.~\ref{figS:loops}. Circles are the integrator of Ref.~\cite{Herderschee2026} run from the code in its arXiv ancillary files, at the 20 of its 34 compared configurations that lie on this line. The two agree to a median 0.2\%.}
\label{fig:S2}
\end{figure}

\begin{figure}[t]
\centering
\includegraphics[width=0.6\textwidth]{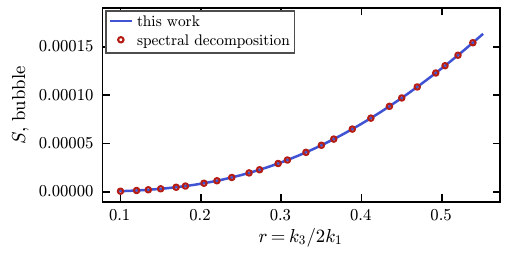}
\caption{Renormalized bubble bispectrum with the loop carrying $\bk_3$, in the model of Ref.~\cite{XianyuZhang2023} at internal $\mu=1.472$ and $k_1=k_2$, from the bubble's residue series (curve) and from the spectral decomposition of Ref.~\cite{XianyuZhang2023} (circles). The two agree to $10^{-6}$ for $r\equiv k_3/2k_1\le0.45$, beyond which the truncation error of the spectral series grows. Here $S=-8k_1k_2k_3^4\,B_{\rm XZ}$ in the notation of Ref.~\cite{XianyuZhang2023}.}
\label{fig:S3}
\end{figure}

\begin{figure}[h]
\centering
\includegraphics[width=0.6\textwidth]{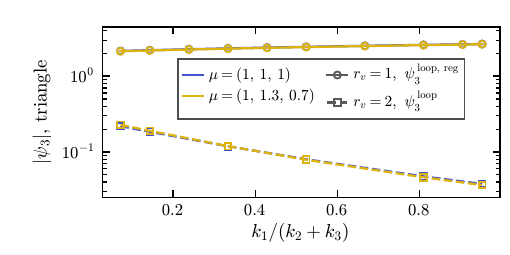}
\caption{The modulus $|\psi_3|$ of the triangle along the family $(k_1,1,1.1)$ at two mass sets $\mu=(\mu_A,\mu_B,\mu_C)$. Solid lines are the cutoff-independent part $\psi_3^{\rm loop,reg}=\psi_3^{\rm loop}-4\pi D_0\ln\Lambda$ of Eq.~(\ref{sm:triangle}) with conformally coupled legs, of the form $-4\pi D_0\ln k_T+b$ fixed by conformal symmetry. Dashed lines are the finite triangle with $\pid$ legs ($r_v=2$). Brute-force integration of Eq.~(1) and the loop momentum (circles and squares) agrees with the curves to better than $4\times10^{-6}$ and $9\times10^{-6}$ respectively.}
\label{figS:loops}
\end{figure}

\textit{Trees.}---Direct Schwinger--Keldysh time integration reproduces the double and triple exchange at 198 isosceles and scalene configurations down to $k_3/k_1=0.002$, including $\mu=2$ on every line. The differences are below $3\times10^{-11}$ at half the default grid step of the line blocks ($2\times10^{-10}$ at the default). The largest, $2.4\times10^{-11}$, is at $\mu=(0.5,1,1.5)$ and $k_3/k_1=0.002$, where $S$ is a near-cancellation of blocks four orders of magnitude larger. Fixed-exponent fits over $10^{-4}\le k_3/k_1\le10^{-2}$ return the coefficients $\cM_2$ and $\cM_3$ to $2\times10^{-10}$ and $5\times10^{-10}$ (with the time grid extended to $\ln t\le16$, as needed below $k_3/k_1\simeq0.002$). Figure~\ref{fig:S2} compares the full correlators with CosmoFlow~\cite{Werth2024}, with couplings mapped through its master action and no free parameter. The median difference is 0.13\% and 0.09\% over 24 configurations each, scalene ones included. The uniform $-0.1\%$ residual is consistent with the $O(\rho^2)\approx4\times10^{-4}$ mixing correction that the transport resums. One exception is near the zero crossings of the double exchange, where relative differences lose meaning and the absolute difference stays below $10^{-3}$ of the equilateral value. A second exception is the points at $k_3/k_1=0.05$, where the 1.4\% deviation is the solver's switch-on transient.

\textit{Bubble.}---The model of Fig.~\ref{fig:S3} is that of Ref.~\cite{XianyuZhang2023}, two derivatively coupled inflaton legs at one vertex and one at the other ($r_v=4$ and 2), in $\overline{\rm MS}$ at renormalization point $\mu_R$ equal to the mass parameter $\mu$. The oscillatory residue series match its spectral series to $10^{-22}$. For $r\equiv k_3/2k_1\le0.45$, where the spectral series is converged, the renormalized bubble matches to $10^{-6}$ once a scheme offset is included. This offset, between minimal subtraction of $2\pi\ln[\Lambda/(x_1{+}x_2)]/(x_1{+}x_2)^3$ and $\overline{\rm MS}$, equals $[\ln(\mu_R/2)+\gamma_E-\half]/4\pi^2$ times the contact shape $(k_3/k_T)^3$.

\textit{Triangle.}---Brute-force integration of the three vertex times and then the loop momentum reproduces Eq.~(\ref{sm:triangle}), logarithm separated, at 33 configurations in shape and mass to a median $3\times10^{-6}$. At fixed masses these brute-force values follow $-4\pi D_0\ln k_T+b$ to $10^{-6}$ with $D_0$ from Eq.~(14) and only $b$ fitted. The Mellin--Barnes values follow the same form to $2\times10^{-6}$. The fit gives $b=0.4358+1.5794\,\ii$ at $\mu=(1,1,1)$ and $0.4336+1.5581\,\ii$ at $(1,1.3,0.7)$. The same integration reproduces the $r_v=2$ triangle at fourteen configurations, seven shapes at each of two mass sets with twelve of them scalene, to a median $5.5\times10^{-6}$, the level of the brute force's quadrature error. Brute-force values at increasing momentum cutoff $\Lambda$ converge as $\Lambda^{-2}$, as expected from the $\ell^{-5}$ tail of the $r_v=2$ integrand, with no logarithm. The integrator of Ref.~\cite{Herderschee2026} (run from its arXiv ancillary code at 34 configurations of its grid, $\mu=0.5$ and 0.7) reproduces our brute-force equilateral-normalized $r_v=2$ triangle to a median 0.2\% and at most 0.6\%, within its quadrature uncertainty. Its unnormalized amplitude reproduces the factor 8 between our couplings and theirs to 0.6\%. Figure~\ref{fig:S2}(c) shows the 20 of these configurations that lie on the line $k_1=k_2$, ten per mass. Assembled into the correlator at $\nu=\half$ with two conformally coupled legs at each vertex ($r_v=2$) and independent vertex energies, the representation reproduces the closed form of Refs.~\cite{Henn2026,Pimentel2026} to $5\times10^{-8}$. At $\mu=(1,1,1)$ and $(k_1,k_2,k_3)=(1.0,1.2,0.9)$, with the bare measure $\dd^3\ell$, $\psi_3^{\rm loop}=-0.05956-0.06447\,\ii$ for $r_v=2$.

In the squeezed limit, with unit dimensionless $\pid\sigma_i\sigma_j$ couplings and the measure $\dd^3\ell/(2\pi)^3$ included, $\cM_{\rm tri}=(1.429-3.719\,\ii)\times10^{-6}$ at $\mu=(1,1,1)$, a fraction $1.3\times10^{-4}$ of the single-exchange coefficient $|\cM_1|$ of Sec.~\ref{sm:double}. Further, $|\cM_{\rm tri}|\,e^{\pi(\mu_B+\mu_C)}$ stays between 1.9 and $2.8\times10^{-3}$ for $\mu_A=1$ and $0.25\le\mu_B,\mu_C\le2$. Fixed-exponent fits of brute-force evaluations of the assembled correlator reproduce it, and the difference-frequency family at unequal masses, to $7\times10^{-5}$ at both mass sets of Fig.~\ref{figS:loops}.

\textit{Scheme constants of the triangle.}---A rescaled regulator $(\ell/\ell_0)^{-2\epsilon}$ in Sec.~\ref{sm:poles} shifts $R$ by $4\pi D_0\ln\ell_0$, and replacing the cutoff $\Lambda=a(\eta_0)H$ on $\ell$ by a cutoff on the vertex times at $\eta_0$ shifts it further by $4\pi(C_D-\ii\pi D_0/2)$. Here $C_D=\int_0^\infty\dd\tau\,\ln\tau\,B(\tau)$, with $B(\tau)$ the integral of Eq.~(14) at fixed earliest vertex time $\tau$, is a complex constant that depends on the masses ($C_D/D_0=0.063+1.839\,\ii$ at $\mu=1$). At equal masses $D_0=\frac18\,\zeta(3,\half+\ii\mu)$ equals $-0.08448-0.01437\,\ii$ at $\mu=1$, where $R=1.634+1.814\,\ii$ at $(k_1,k_2,k_3)=(0.9,1,1.1)$.

Figure~\ref{figS:loops} compares the triangle of Eq.~(\ref{sm:triangle}) with brute-force integration.

\section{Closed forms at physical kinematics}\label{sm:closed}

Closing the contours of Eq.~(3) in a fixed order gives closed forms that converge at every physical shape. We first integrate over the Mellin variables that the energies do not enter. In a tree these are the variables at the mixing ends, where the energies of the line and the leg coincide. In the bubble they are the two directions at fixed vertex sums. What remains then closes in series whose argument stays below one.

\textit{Dressed legs.}---At a mixing vertex $x_l=y_l=k_l$, so the line power $(y_l/2)^{-2\delta_l}$ and the vertex factor of Eq.~(3) combine into $k_l^{\nu_l+2s_l'-1/2}$ times a function of $s_l$ alone. The $s_l$ integral is therefore the mixing time integral done in closed form. In the correlator it gives the dressed leg $w_l=\F_l-\frac1\pi\hat\Phi_lK_{\ii\mu_l}$ of Sec.~\ref{sm:double}. Following Ref.~\cite{Pinol2026strong} (see also the Laplace-space representation of Refs.~\cite{Belrhali2026laplace,Belrhali2026strong}), we write it as a Laplace transform of Legendre functions,
\begin{equation}
w_l(t)=2t^{1/2}e^{-t}\int_0^\infty\dd x\,e^{-2tx}f_l(x)\,,\qquad f_l(x)=Q_{d_l-1}(1+2x)+c^+_lP_{d_l-1}(1+2x)\,, \label{sm:leg}
\end{equation}
with $d_l=\half-\ii\mu_l$ and $c^\pm_l=-\frac\pi2\,(1\pm\ii e^{\pm\pi\mu_l})/\cosh\pi\mu_l$. The Legendre functions are those of argument above one, $P_{d-1}(1+2x)={}_2F_1(d,1-d;1;-x)$ and $Q_{d-1}(1+2x)=\frac{\Gamma(d)^2}{2\Gamma(2d)}(1+x)^{-d}\,{}_2F_1\bigl(d,d;2d;\frac1{1+x}\bigr)$. We confirm Eq.~(\ref{sm:leg}) pointwise to $10^{-25}$. Each leg then enters through the one-leg function
\begin{equation}
\begin{aligned}
\Psi_g(\rho)=\int_0^\infty\!\dd x\,\frac{f(x)}{(1+\rho x)^g}={}&\frac{{}_3F_2(g,1,1;2-d,1+d;\rho)}{2d(d-1)}+B_g\,\rho^{d-1}\,{}_2F_1(g+d-1,d;2d;\rho)\\
&+C_g\,\rho^{-d}\,{}_2F_1(g-d,1-d;2-2d;\rho)\,,
\end{aligned}\label{sm:Psi}
\end{equation}
with $B_g=\frac{\Gamma(g+d-1)}{\Gamma(g)}\bigl[\frac{\Gamma(1-d)\Gamma(d)^2}{2\Gamma(2d)}+c^+\frac{\Gamma(1-2d)}{\Gamma(1-d)}\bigr]$ and $C_g=c^+\,\frac{\Gamma(g-d)\Gamma(2d-1)}{\Gamma(g)\Gamma(d)}$. We obtain it by closing a Mellin--Barnes integral built on the Mellin transforms of $Q_{d-1}(1+2x)$ and $P_{d-1}(1+2x)$, which are $\Gamma(a)^2\Gamma(d-a)/[2\Gamma(d+a)]$ and $\Gamma(a)\Gamma(1-d-a)\Gamma(d-a)/[\Gamma(1-d)\Gamma(d)\Gamma(1-a)]$. The three terms are the analytic background and the two signal families. They converge for $\rho<1$, and Eq.~(\ref{sm:Psi}) matches quadrature to $10^{-27}$.

\textit{Triple exchange.}---With the legs of Eq.~(\ref{sm:leg}) the central time integral is elementary. For the leg assignment of Fig.~2 we find
\begin{equation}
S=-\frac{k_1k_2k_3}{2}\,\Re\int_1^\infty\!\!\int_1^\infty\!\!\int_1^\infty\!\dd^3z\;\frac{F_A(z_1)\,F_B(z_2)\,F_C(z_3)}{(k_1z_1+k_2z_2+k_3z_3)^3}\,,\qquad F_l(z)=f_l\bigl(\tfrac{z-1}{2}\bigr)\,. \label{sm:master}
\end{equation}
The masses sit in the densities and the momenta in the kernel. With $u_l=2k_l/k_T$ and $z_l=1+2x_l$ the kernel is proportional to $(D+u_3x_3)^{-3}$, where $D=1+u_1x_1+u_2x_2\ge1$. The leg on $\bk_3$ then gives $D^{-3}\Psi^{(C)}_3(u_3/D)=\sum_ea_e\,u_3^{\,e}D^{-3-e}$, where $e$ runs over the families $n$, $d_C-1+m$, and $m-d_C$ ($n,m\ge0$). Its argument $u_3/D$ never exceeds $u_3$. The other two legs leave $\iint\dd x_1\dd x_2\,f_Af_B\,D^{-s}$ with $s=3+e$. With $t_i=u_ix_i/(1+u_ix_i)$ this kernel factorizes, since $D=(1-t_1t_2)/[(1-t_1)(1-t_2)]$, and expanding $(1-t_1t_2)^{-s}$ gives
\begin{gather}
S=-\frac{u_3}{2}\,\Re\sum_e a_e\,u_3^{\,e}\,\mathcal{J}(3+e)\,,\qquad \mathcal{J}(s)=\sum_{k\ge0}\frac{(s)_k}{k!}\,W^A_k\bigl[u_1^{-1}\bigr](s)\,W^B_k\bigl[u_2^{-1}\bigr](s)\,,\label{sm:triple}\\
W_k[c](s)=\int_0^1\!\dd t\;t^k(1-t)^{s-2}f\Bigl(\frac{ct}{1-t}\Bigr)=\sum_{m\ge0}\gamma_m\,B(k+1,s-1+\alpha_m)\;{}_2F_1(\alpha_m,k+1;s+\alpha_m+k;1-c)\,.\nonumber
\end{gather}
Here $\alpha_m$ runs over $d+m$ and $1-d+m$, and $\gamma_m$ are the coefficients of $f$ in powers of $(1+x)^{-1}$, namely $\bigl[\frac{\Gamma(d)^2}{2\Gamma(2d)}+c^+\frac{\Gamma(1-2d)}{\Gamma(1-d)^2}\bigr]\frac{(d)_m^2}{(2d)_m\,m!}$ and $c^+\frac{\Gamma(2d-1)}{\Gamma(d)^2}\frac{(1-d)_m^2}{(2-2d)_m\,m!}$. The diagram is symmetric under permutations of the pairs $(k_l,\mu_l)$. Expanding the leg with the smallest momentum gives $u_3/D\le u_3\le\frac23$, inside the radius of convergence of Eq.~(\ref{sm:Psi}), so Eq.~(\ref{sm:triple}) converges at every physical shape, folded ones included. It reproduces Eq.~(3) at $\mu=(1,1.5,0.7)$ to between $1.5\times10^{-10}$ and $3\times10^{-8}$ at six shapes from $(1,1,0.02)$ to the folded $(1,0.5,0.5)$, limited by the double-precision moments of our implementation. The closed form of $W_k$ matches quadrature to between $10^{-9}$ and $10^{-21}$, and at small $k$ its $m$ sum needs acceleration. Expanding the hard legs as well reproduces the series of Ref.~\cite{XianyuZang2026}, which needs $u_1+u_2+u_3<1$. The same steps give the single and double exchange, whose full shapes are known~\cite{QinXianyu2023closed,Aoki2024}, as one ${}_3F_2$ and two ${}_2F_1$, and as one sum of the functions of Eq.~(\ref{sm:Psi}). Both match Eq.~(3) to $10^{-11}$.

\textit{The squeezed coefficient at general masses.}---At $u_1=u_2=1$ the moments $W_k$ are single ${}_3F_2$ at unit argument. With $W_k=Q_k+c\,P_k$,
\begin{equation}
Q_k=\frac{\Gamma(d)^2\,(k!)^2}{2\,\Gamma(k+1+d)^2}\;{}_3F_2\!\left(\begin{matrix}1+d-s,\;k+1,\;k+1\\ k+1+d,\;k+1+d\end{matrix};1\right),\qquad
P_k=B(k+1,\,s+d-1)\;{}_3F_2\!\left(\begin{matrix}d,\;d,\;k+1\\ 1,\;s+d+k\end{matrix};1\right).
\end{equation}
Since $\overline{Q_{d-1}}=Q_{d-1}-\pi\cot(\pi d)\,P_{d-1}$, both terms of Eq.~(\ref{sm:M3mom}) carry the same Legendre functions. At $\bar s=\frac52-\ii\mu_C$ we find
\begin{equation}
\cM_3=-\frac{\hat\Phi_C\,2^{-1/2-2\ii\mu_C}}{8\,\Gamma(1+\ii\mu_C)}\;\overline{\Bigl[\bar\lambda_+\,\mathcal{J}_1(\bar s;c^-_A,c^-_B)+\bar\lambda_-\,\mathcal{J}_1(\bar s;c^+_A,c^+_B)\Bigr]}\,,\qquad \mathcal{J}_1(s;c_A,c_B)=\Gamma(s)\sum_{k\ge0}\frac{(s)_k}{k!}\,W^A_k\,W^B_k\,,\label{sm:M3}
\end{equation}
with $W^l_k=Q^l_k+c_lP^l_k$ and $\lambda_\pm$ at $\mu_C$. The sum over $k$ converges as $k^{-5/2}$. Equation~(\ref{sm:M3}) reproduces Eq.~(\ref{sm:M3mom}) to $4\times10^{-11}$ at six mass sets, four of them unequal. At equal masses the ${}_3F_2$ in $Q_k$ terminates and the mixed terms vanish. The result agrees with the closed form of Ref.~\cite{Pinol2026strong}, $A_3e^{\ii\delta_3}/\lambda^3=64\ii\,\overline{\cM_3}$ in the notation of its Eq.~(6.17) with $\mathsf P(d)$ evaluated from its Eq.~(E.3), to $4\times10^{-10}$ or better at $\mu=0.5$, $0.8$, $1$, $1.5$, $2$, $3$, and $4$.

\textit{The bubble.}---In $\psi_2^{\rm MB}$ the energies enter only through the vertex sums $\sigma_v=s_{v,1}+s_{v,2}$, which fix $\Sigma_v=\bar\nu+2\sigma_v$ and $\cS=-\bar\nu-\sigma_1-\sigma_2$, with $\bar\nu=\nu_1+\nu_2$. Integrating over the two remaining directions at fixed $\sigma_v$ gives
\begin{equation}
2^{2\cS}C_2\iint\!\frac{\dd s_{1,1}\,\dd s_{2,1}}{(2\pi\ii)^2}\prod_{j=1,2}\frac{\Gamma(-s_{1,j})\Gamma(-s_{2,j})\Gamma(\delta_j)\Gamma(1-\delta_j)}{2\,\Gamma(\nu_j{+}s_{1,j}{+}1)\Gamma(\nu_j{+}s_{2,j}{+}1)}
=\sum_{n\ge0}8\pi G_n\,\frac{\Gamma(n-\sigma_1)\Gamma(n-\sigma_2)\,\Gamma(\delta')\Gamma(1-\delta')\,2^{2\delta'}}{2\,\Gamma(\nu'_n+\sigma_1-n+1)\,\Gamma(\nu'_n+\sigma_2-n+1)}\,,\label{sm:rel}
\end{equation}
with $\nu'_n=\bar\nu+\thalf+2n$, $\delta'=\cS-\thalf$, and
\begin{equation}
G_n=\frac{\Gamma(\thalf+n)\,\Gamma(\thalf+\nu_1+n)\,\Gamma(\thalf+\nu_2+n)\,\Gamma(\thalf+\bar\nu+n)\,(\thalf+\bar\nu+2n)\,\Gamma(1+\bar\nu+2n)}{n!\,\Gamma(1+\nu_1+n)\,\Gamma(1+\nu_2+n)\,\Gamma(1+\bar\nu+n)\,\Gamma(3+\bar\nu+2n)}\,.
\end{equation}
We confirm Eq.~(\ref{sm:rel}) to $10^{-10}$ or better at four complex points with $\half<\Re\cS<\thalf$. Its right side is the Mellin form of the product identity of bulk-to-bulk propagators~\cite{FitzpatrickKaplan2012a,FitzpatrickKaplan2012b,SleightTaronna2021b}, the integrand of a single exchange of order $\nu'_n$ with $s_v=\sigma_v-n$. On the contours of $\psi_2^{\rm MB}$ each term has $-1<\Re\delta'<0$, one pole of $\Gamma(\delta')$ outside the window of Eq.~(2). Moving it back crosses only $\delta'=0$, whose residue is the same tree at zero momentum through the line. Hence
\begin{equation}
\psi_2^{\rm MB}=8\pi\sum_{n\ge0}G_n\Bigl[\psi_1^{(\nu'_n)}(x_1,x_2,k)-\psi_1^{(\nu'_n)}(x_1,x_2,0)\Bigr]\,,\label{sm:tower}
\end{equation}
where $\psi_1^{(\nu)}$ is the single exchange of order $\nu$ through momentum $k$ at vertex powers $a_v=r_v-\thalf$. The subtracted trees are elementary,
\begin{equation}
\psi_1^{(\nu)}(x_1,x_2,0)=\frac{\Gamma(A)}{2\nu\,E^{A}}\left[\frac{{}_2F_1(1,A;a_2+\nu+1;x_2/E)}{a_2+\nu}+\frac{{}_2F_1(1,A;a_1+\nu+1;x_1/E)}{a_1+\nu}\right],
\end{equation}
with $A=a_1+a_2$ and $E=x_1+x_2$. The terms of Eq.~(\ref{sm:tower}) fall as $n^{-3}$. Each term matches its own contour integral to $10^{-11}$ for $n=0$, 1, and 3 at two shapes. In the bispectrum with the loop carrying $\bk_3$ and the single leg at the second vertex, $x_2=k=k_3$. At the vertex powers $(4,2)$ of Ref.~\cite{XianyuZhang2023} each tree is then
\begin{equation}
\psi_1^{(\nu)}(k_1{+}k_2,k_3,k_3)=\frac{4}{k_T^3}\,\Psi^Q_3(\rho)\,,\qquad \rho=\frac{2k_3}{k_T}\,,\quad d=\nu+\half\,,
\end{equation}
with $\Psi^Q$ the $Q$ part of Eq.~(\ref{sm:Psi}), one ${}_3F_2$ and one ${}_2F_1$. The sum then converges at every shape short of the folded limit. It matches an independent one-dimensional integral over the invariant distance between the two vertices to $2\times10^{-12}$ at five shapes up to $\rho=0.98$.

\textit{The in-in bubble.}---By Eq.~(\ref{sm:eq15}), the Feynman kernel is $(e^{\pi\mu}\G^+-e^{-\pi\mu}\G^-)/(2\sinh\pi\mu)$, with $\G^\pm=K_\nu(yt_>)I_{\pm\nu}(yt_<)$, and the cut kernel is $\frac1\pi K_\nu(yt)K_\nu(yt')=(\G^--\G^+)/(2\sin\pi\nu)$. For real vertex phases the sum over branches of the bubble is therefore $\sum_\epsilon W_\epsilon\,\G^{\epsilon_1}_1\G^{\epsilon_2}_2$ at fixed loop momentum, with
\begin{equation}
W_\epsilon=\frac{\epsilon_1\epsilon_2\,\sinh^2\bigl[\frac\pi2(\epsilon_1\mu_1+\epsilon_2\mu_2)\bigr]}{\sinh\pi\mu_1\,\sinh\pi\mu_2}\,.
\end{equation}
The shadow $\G^-$ is the kernel at order $-\nu$, so each sign pattern is Eq.~(\ref{sm:tower}) at orders $(\epsilon_1\nu_1,\epsilon_2\nu_2)$. The patterns with $\epsilon_1\epsilon_2=-1$ carry the difference frequency. At fixed line energies this decomposition reproduces the in-in integrand of a separate time-domain code by the author to $4\times10^{-16}$. In the two-mass version of the model of Ref.~\cite{XianyuZhang2023}, the assembled bubble bispectrum matches that code, integrated over the loop momentum with a cutoff, at $\mu=(1.2,2)$ and six shapes, $(1,1,1)$, $(1,1,0.3)$, $(1,0.8,0.5)$, $(0.6,1,1.2)$, $(1,1,1.9)$, and $(1,1,0.05)$, to $3\times10^{-12}$ after one constant common to all shapes. This constant multiplies the shape $k_1k_2k_3/k_T^3$ of the logarithm, so it only fixes the renormalization scale.

\textit{The triangle.}---The vertex energies of the triangle are its external momenta, so its kinematics enter through two ratios, and six of its eight Mellin variables are free of them. No pair of its lines shares both end points, however, and the product identity does not apply. The integral over those six variables is the de~Sitter counterpart of the anti--de~Sitter triangle, which Ref.~\cite{Liu2018sixj} identifies with a $6j$ symbol of the conformal group. That object is known in closed form for $d=1$, 2, and 4 but not for $d=3$, so we keep the triangle as the finite integral of Eq.~(3) with $C_3$ inserted.

\textit{The squeezed background of the triple exchange.}---The same sum at $s=3$ in place of $\bar s$ gives the non-oscillatory squeezed term of the triple exchange. In the normalization of Ref.~\cite{XianyuZang2026} [their Eq.~(110)] it gives $\mu^6C_B^{(3)}=-0.5704$ at $\mu=1$ and $0.3597$ at $\mu=2$ at equal masses. It changes sign near $\mu\simeq1.83$, as Ref.~\cite{XianyuZang2026} observes, and tends to one at large mass ($1.04$ at $\mu=12$).

\textit{Relation to the series of Ref.~\cite{XianyuZang2026}.}---The dimensionless mixed propagator of Ref.~\cite{XianyuZang2026} [their Eq.~(106)] on the rotated axis is $-\frac12z^{3/2}w(z)$, with $w$ the dressed leg of Eq.~(\ref{sm:leg}), to 25 digits at $\mu=1$ and 2. Their squeezed coefficients, which they evaluate by quadrature, are therefore the moments above, and their signal coefficient is $4^{1/2+\ii\mu}\,16\,\cM_3$ at equal masses, to $10^{-12}$. Their full star diagram [their Eq.~(85)] is a sum of 27 trivariate series in $u_i=2K_i/(E_0+K_{123})$ that converges only for $u_1+u_2+u_3<1$, while the bispectrum sits at $u_1+u_2+u_3=2$. Inside the convergence region it equals a one-dimensional integral over three dressed legs to 12 digits at four kinematic points, and that integral converges at every physical $u_i$. At $E_0=0$ it reproduces the triple-exchange correlator of Eq.~(3) to $10^{-11}$ at four shapes. The triple exchange of the main text is therefore the analytic continuation of their series to physical kinematics. We also note that the prefactor $64$ of their Eq.~(104) should read $128$, which is what their series definition, Eq.~(103), gives (we find the ratio 2 to 20 digits). With $128$ their large-mass limits, Eq.~(118), hold.

\textit{Logarithm and correlator constant of the conformally coupled triangle.}---With one conformally coupled leg at each vertex the correlator is a three-point function of dimension-one operators, which has no anomaly, so conformal symmetry fixes it to one constant times $1/(k_1k_2k_3)$. Up to the couplings and leg factors stripped in the main text, $\langle\phi\phi\phi\rangle'\propto\eta_0^3\,\mathfrak{B}/(k_1k_2k_3)$, with $\mathfrak{B}=\int\dd^3\ell\,J$. Here $J$ is the loop integrand of the correlator assembled with the cutting rule of Sec.~S3 (the uncut term and the three terms in which the two lines at one vertex are cut). Every kernel of the correlator is a combination of $\G_{\pm\nu}$, through $G^F=(e^{\pi\mu}\G_\nu-e^{-\pi\mu}\G_{-\nu})/(2\sinh\pi\mu)$, $K_\nu(yt)K_\nu(yt')=\pi(\G_{-\nu}-\G_\nu)/(2\ii\sinh\pi\mu)$, and $\overline{\G_\nu}=\G_{-\nu}$. The integrand is therefore a combination of the eight wavefunction triangles $T_\sigma$ of orders $(\sigma_A\nu_A,\sigma_B\nu_B,\sigma_C\nu_C)$ with real weights
\begin{equation}
w_\sigma=\frac{\sigma_A\sigma_B\sigma_C}{8\sinh\pi\mu_A\sinh\pi\mu_B\sinh\pi\mu_C}\Bigl[e^{\pi(\sigma_A\mu_A+\sigma_B\mu_B+\sigma_C\mu_C)}+\sum_le^{-\sigma_l\pi\mu_l}\Bigr]\,. \label{sm:wsigma}
\end{equation}
Each $T_\sigma$ has the conformal form $-4\pi D_0(\sigma\nu)\ln k_T+b_\sigma$, so the coefficient of the logarithm in the correlator is $\Im Q$ with $Q=\sum_\sigma w_\sigma D_0(\sigma\nu)$, Eq.~(14) at sign-flipped orders. This reproduces to $3\times10^{-11}$ the numerical tail constant $D_0^F-\sum_vc_v$, with $D_0^F$ and $c_v$ the unit-energy constants of the uncut block and of the three blocks with the two lines at vertex $v$ cut. At unequal masses $D_0$ is a sum of closed-form terms. With the digamma function $\psi$, let $\beta(z)=\frac12[\psi(\frac{z+1}{2})-\psi(\frac z2)]$, $b_l=\beta(\nu_l+n+\frac12)$, $\tilde b_l=\beta(\nu_l+\frac12-n)$, and $\mathsf{B}(x,y)=\sum_{m\ge0}(-1)^m\beta(x+m)\beta(y+m)$. Then
\begin{equation}
\begin{aligned}
2D_0={}&\sum_{n\ge0}\Big[-b_Ab_Bb_C+\tilde b_Ab_Bb_C+b_A\tilde b_Bb_C+b_Ab_B\tilde b_C\Big]-\prod_l\beta(\nu_l+\tfrac12)\\
&+\sum_X\pi\tan(\pi\nu_X)\,\mathsf{B}(1{+}\nu_X{+}\nu_Y,\,1{+}\nu_X{+}\nu_Z)\,,
\end{aligned}\label{sm:D0unequal}
\end{equation}
where $X$ runs over the three lines, with $Y$ and $Z$ the other two. Equation~(\ref{sm:D0unequal}) reproduces $\frac18\zeta(3,\half+\ii\mu)$ at equal masses to 30 digits. Direct quadrature of Eq.~(14) reproduces it to $10^{-9}$ or better at seven mass sets, including three light lines ($\nu=0.3$, 0.4, and 0.2), mixed light and heavy lines, and a very light line next to a heavy one. In this quadrature the times beyond a cutoff $T$ contribute $3/(32T^2)$ at leading order, independently of the masses, and we fit the next two orders in $1/T$. With Eq.~(\ref{sm:D0unequal}) at 30-digit precision, $|\Im Q|/|Q|$ is below $10^{-30}$ at $\mu=(1,1,1)$, $(1,1.3,0.7)$, and $(0.6,1.5,2.2)$, and at equal masses $Q=\Re D_0(\nu,\nu,\nu)$. The cancellation also follows from unitarity. At the late times $t_v\sim1/\ell$ that generate the $\ell^{-3}$ tail of the loop integrand, the legs act as equal sources on the two branches of the in-in contour, so the branch sum of the tail vanishes. The constant is $\mathfrak{B}=-2\Im\sum_\sigma w_\sigma b_\sigma$, in which the scheme dependence of the $b_\sigma$ cancels because $Q$ is real. Since $\mathfrak{B}$ does not depend on the momenta, we evaluate this loop integral directly at $k_3\to0$ and find $\mathfrak{B}=-3.11475$ at $\mu=(1,1,1)$ and $-3.02036$ at $(1,1.3,0.7)$. The three assignments of the unequal masses to the lines agree to $10^{-6}$, and an independent collinear configuration agrees to $2\times10^{-5}$.

The Lorentzian phase of the late-time logarithm gives the leading part. The vertex times are cut at $t_0=\ii|\eta_0|$, so $\ln(1/t_0)$ carries the phase $-\ii\pi/2$, and in the weighted sum this contributes $4\pi^2Q$. Hence $\mathfrak{B}=4\pi^2Q+\Delta$, with $4\pi^2Q=\frac{\pi^2}{2}\Re\,\zeta(3,\half+\ii\mu)$ at equal masses. The remainder is Boltzmann suppressed, $\Delta=-0.646$, $0.220$, and $6.5\times10^{-4}$ at $\mu=0.5$, 1, and 2, and consistent with zero at $\mu=3$. For heavy lines the triangle therefore acts as a local cubic vertex whose only effect on the correlator is the phase of its secular logarithm. Finally, since $\psi_3+4\pi D_0\ln k_T=b$ holds at all positive energies, we may send two vertex energies to zero. Both vertex factors then close on their delta functions, $C_3$ becomes a ratio of Gamma functions, and $b$ and $\mathfrak{B}$ reduce to a four-fold Mellin--Barnes integral of Gamma functions plus the derivative at $\cS=\thalf$ of a three-fold one. This form reproduces the values above ($\mathfrak{B}=-3.1147521$ at $\mu=(1,1,1)$) and $b(1,1,1)=0.4358371+1.5793549\,\ii$. We have not closed these integrals, whose inner integrals are of ${}_4F_3$ type.

\textit{An expansion of $C_3$ at physical momenta.}---Closing $u$ and $u'$ in Eq.~(13) to the right gives the four Appell $F_4$ series of Refs.~\cite{BoosDavydychev1991,Davydychev1992}, which converge only for $\sqrt w+\sqrt z<1$~\cite{Exton1995,Ananthanarayan2020}. Physical momenta have $\sqrt w+\sqrt z\ge1$, so we use Eq.~(13) directly. In its convergent domain $C_3$ is the triple-$K$ integral of Refs.~\cite{Bzowski2014,Coriano2013},
\begin{equation}
C_3=\frac{2^{5/2}\pi^{3/2}}{\Gamma(3-\cS)\,\Gamma(\delta_A)\Gamma(\delta_B)\Gamma(\delta_C)}\int_0^\infty\!\dd x\;x^{1/2}\prod_{j=1}^3k_j^{\beta_j}K_{\beta_j}(k_jx)\,,
\end{equation}
with $\beta_1=\frac32-\delta_A-\delta_C$, $\beta_2=\frac32-\delta_A-\delta_B$, and $\beta_3=\frac32-\delta_B-\delta_C$. We split $K_{\beta_3}(k_3x)$ into its two ascending series and integrate term by term. Each term is a moment of two Bessel functions, given by a Gauss function of $1-k_2^2/k_1^2$. With $\lambda=\cS-\frac32$ and $\delta_{BC}=\delta_B+\delta_C$, the two series give
\begin{multline}
C_3=\pi^{3/2}\,k_2^{3-2\delta_A-2\delta_B}\Bigl[C_{\rm an}\,k_1^{-2\delta_C}\;\Xi\bigl(\tfrac32{-}\delta_A,\delta_C;\lambda,\delta_B;\delta_{BC}{-}\tfrac12,\delta_{BC}\bigr)\\
+C_{\rm nl}\,k_1^{2\delta_B-3}\,k_3^{3-2\delta_{BC}}\;\Xi\bigl(3{-}\cS,\tfrac32{-}\delta_B;\delta_A,\tfrac32{-}\delta_C;\tfrac52{-}\delta_{BC},3{-}\delta_{BC}\bigr)\Bigr]\,, \label{sm:C3exp}
\end{multline}
where
\begin{equation}
C_{\rm an}=\frac{\Gamma(\lambda)\Gamma(\frac32-\delta_A)\Gamma(\frac32-\delta_{BC})}{\Gamma(\delta_A)\Gamma(\delta_{BC})\Gamma(3-\cS)}\,,\qquad
C_{\rm nl}=\frac{\Gamma(\delta_{BC}-\frac32)\Gamma(\frac32-\delta_B)\Gamma(\frac32-\delta_C)}{\Gamma(\delta_B)\Gamma(\delta_C)\Gamma(3-\delta_{BC})}\,,
\end{equation}
and $\Xi$ is the double series of Srivastava--Daoust type~\cite{SrivastavaDaoust1969}
\begin{equation}
\Xi(a,b;a',b';c,e)=\sum_{n,m\ge0}\frac{(a)_{n+m}(b)_{n+m}(a')_n(b')_n}{(c)_n\,(e)_{2n+m}}\,\frac{X^nY^m}{n!\,m!}\,,\qquad X=\frac{k_3^2}{k_1^2}\,,\quad Y=1-\frac{k_2^2}{k_1^2}\,,
\end{equation}
whose sum over $m$ at fixed $n$ is a Gauss function. The first term is analytic in $k_3^2$ and carries $\Gamma(\lambda)$, and the second carries the non-analytic power $k_3^{3-2\delta_{BC}}$. With the momenta labeled so that $k_3\le k_2\le k_1$, the terms of the series in $X$ fall at least as $4^{-n}$, so Eq.~(\ref{sm:C3exp}) converges at every physical shape and, by continuation, on the tube $0<\Re\delta_l<\frac32$. On the isosceles slice $k_1=k_2=k$ one has $Y=0$, and each double series becomes a ${}_4F_3$ of $\xi=k_3^2/(4k^2)$,
\begin{equation}
C_3=\pi^{3/2}k^{3-2\cS}\biggl[C_{\rm an}\,{}_4F_3\biggl(\begin{matrix}\lambda,\ \delta_B,\ \delta_C,\ \frac32{-}\delta_A\\ \delta_{BC}{-}\frac12,\ \frac{\delta_{BC}}{2},\ \frac{\delta_{BC}+1}{2}\end{matrix};\xi\biggr)+C_{\rm nl}\Bigl(\frac{k_3}{k}\Bigr)^{3-2\delta_{BC}}{}_4F_3\biggl(\begin{matrix}\delta_A,\ 3{-}\cS,\ \frac32{-}\delta_B,\ \frac32{-}\delta_C\\ \frac52{-}\delta_{BC},\ \frac{3-\delta_{BC}}{2},\ \frac{4-\delta_{BC}}{2}\end{matrix};\xi\biggr)\biggr]\,. \label{sm:C3iso}
\end{equation}
Equation~(\ref{sm:C3exp}) reproduces the Schwinger-parameter form of $C_3$ at three shapes, for real and for complex exponents, to between $6\times10^{-9}$ and $10^{-6}$, the accuracy of our two-dimensional quadrature. On the isosceles slice Eq.~(\ref{sm:C3iso}) reproduces it to machine precision.

\textit{The triangle with $\pid$ legs to all orders in the squeezed-limit expansion.}---The triangle does not close at general kinematics, but its collider signal does. The expansion of Eq.~(\ref{sm:C3exp}) in powers of $X=k_3^2/k_1^2$, which converges at every physical shape, can be inserted order by order into Eq.~(3) for this triangle ($r_v=2$). The resulting series converges geometrically, with ratio close to $X/4$, and reproduces the contraction with the full $C_3$ to $3\times10^{-12}$. Each order replaces $C_3$ by Gamma functions and Pochhammer symbols of the $\delta_l$. Squeeze $\bk_3$ on the vertex joining $B$ and $C$, at $k_1=k_2=k$. A hard loop momentum gives terms analytic in $k_3$. A soft loop momentum gives the non-analytic terms. Among these only a soft third vertex time contributes, with the soft lines reaching the hard vertices through $I_\nu$ (with a hard third vertex every piece is a scaleless integral). Hence
\begin{equation}
\psi_3=\sum_{N\ge0}a_N\,k_3^N+\frac1{k^3}\Big(\frac{k_3}{k}\Big)^{1+\nu_B+\nu_C}\sum_{N\ge0}c_N\Big(\frac{k_3}{k}\Big)^{2N}\,, \label{sm:psisq}
\end{equation}
with no other non-analytic family. The pair family comes from the poles of $\Gamma(-s)$ at the far ends of the two soft lines. The integral over their near ends then closes by two first Barnes lemmas, and in the analytic continuation a selection rule removes every term but one,
\begin{multline}
c_N=\frac{\pi^2}{2^{1+\nu_{BC}}}\,\frac{\Gamma(-1-\nu_{BC}-2N)}{2+\nu_{BC}+2N}\sum_{i+j+n=N}\frac{4^{-i-j}\,\Gamma(\frac32+\nu_B+2j+n)\,\Gamma(\frac32+\nu_C+2i+n)}{i!\,j!\,n!\,\Gamma(1+\nu_B+j)\,\Gamma(1+\nu_C+i)}\\
\times2\,\big(-\tfrac12y\partial_y\big)_n\,\mathcal{H}_{ij}(y)\Big|_{y=1}\,, \label{sm:cN}
\end{multline}
with $\nu_{BC}=\nu_B+\nu_C$, $(D)_n=D(D+1)\cdots(D+n-1)$, and the hard factor
\begin{equation}
\mathcal{H}_{ij}(y)=\int_0^\infty\!\dd t_1\dd t_2\;t_1^{1+\nu_C+2i}\,t_2^{1+\nu_B+2j}\,e^{-t_1-t_2}\,K_{\nu_A}(yt_>)\,I_{\nu_A}(yt_<)\,.
\end{equation}
The derivatives at $y=1$ are finite combinations of the doubly folded seed of End Matter~C at integer-shifted exponents, each a ratio of Gamma functions times one ${}_3F_2$ at unit argument. Equation~(\ref{sm:cN}) is therefore a finite double sum of such terms at every order, and at $N=0$ it gives $\cM_{\rm tri}$. The selection rule comes from a factor $1/\Gamma(r_3-1-J-n+K+q)$ left by the second lemma. Here $r_3$ is the power at the soft vertex, $J=i+j$, and $K\le J$ and $q\le n$ label the terms of the polynomial expansions. At $r_3=2$ only $K=J$ and $q=n$ survive. At $r_3=1$ the factor vanishes for every term, so the conformally coupled triangle has no soft-limit oscillation at any order. The ratio $|c_{N+1}/c_N|$ falls from $0.52$ at $N=0$ to $0.267$ at $N=13$ (at $\mu=(1,1.3,0.7)$) and approaches $\frac14$, so the pair series converges for $k_3<2k$.

Every kernel of the assembled correlator is a combination of the ordered kernels at $\pm\nu$, $G^F=(e^{\pi\mu}\G_\nu-e^{-\pi\mu}\G_{-\nu})/(2\sinh\pi\mu)$ and $K_\nu K_\nu/\pi=(\G_{-\nu}-\G_\nu)/(2\ii\sinh\pi\mu)$. The structures $FFF$, $FXX$, $XFX$, and $XXF$ then give the real weights $w_\sigma$ of Eq.~(\ref{sm:wsigma}) on the eight wavefunction triangles at orders $\sigma\nu$, and the non-analytic part of the bispectrum is
\begin{equation}
S\supset2\Re\sum_{N\ge0}\Big[\cM^{(N)}_{+}\Big(\frac{k_3}{k}\Big)^{2+\ii(\mu_B+\mu_C)+2N}+\cM^{(N)}_{-}\Big(\frac{k_3}{k}\Big)^{2+\ii(\mu_B-\mu_C)+2N}\Big]\,, \label{sm:Ssq}
\end{equation}
where, with unit dimensionless couplings and the measure $\dd^3\ell/(2\pi)^3$ as in Sec.~\ref{sm:checks},
\begin{equation}
\cM^{(N)}_\pm=-\frac{1}{8(2\pi)^3}\sum_{\sigma_A=\pm}\Bigl[w_{(\sigma_A,+,\pm)}\,c_N(\sigma_A\nu_A,\nu_B,\pm\nu_C)+\bigl(w_{(\sigma_A,-,\mp)}\,c_N(\sigma_A\nu_A,-\nu_B,\mp\nu_C)\bigr)^*\Bigr]\,, \label{sm:Mpm}
\end{equation}
a sum of four terms, with $c_N(\nu_A,\nu_B,\nu_C)$ that of Eq.~(\ref{sm:cN}). At $N=0$, $\cM^{(0)}_+$ reproduces $\cM_{\rm tri}$ to $10^{-21}$ at $\mu=(1,1,1)$, $(1,1.3,0.7)$, and $(0.8,1.6,1.2)$. Equation~(\ref{sm:Ssq}) is the complete collider signal of this triangle to all orders in $k_3/k$ along $k_1=k_2$, both frequencies included. At $\mu=(1,1.3,0.7)$ the first amplitudes are
\begin{center}
\begin{tabular}{ccc}
$N$ & $10^6\,\cM^{(N)}_+$ & $10^6\,\cM^{(N)}_-$\\
\hline
0 & $1.408384-3.737436\,\ii$ & $-2.508212-1.824692\,\ii$\\
1 & $1.203152-1.688839\,\ii$ & $-1.227951-1.047459\,\ii$\\
2 & $0.550621-0.557698\,\ii$ & $-0.443801-0.421439\,\ii$\\
3 & $0.202872-0.164167\,\ii$ & $-0.142357-0.146279\,\ii$\\
4 & $0.067167-0.045654\,\ii$ & $-0.042901-0.046871\,\ii$
\end{tabular}
\end{center}
and the ratios approach $\frac14$. The first correction is therefore about half the leading term at $k_3=k$, and this holds at other masses. Over eleven mass sets with $0.5\le\mu_l\le3$ (four with three equal masses, seven unequal), $|\cM^{(1)}/\cM^{(0)}|$ lies between $0.50$ and $0.63$ for the pair family and, on the unequal sets, between $0.47$ and $0.59$ for the difference family. The ratio $|\cM^{(2)}/\cM^{(1)}|$ lies between $0.36$ and $0.42$. At $\mu_B=\mu_C$ the difference family enters Eq.~(\ref{sm:cN}) at $\nu_B-\nu_C=0$, where $\Gamma(-1-2N)$ has a pole. The pole cancels in Eq.~(\ref{sm:Mpm}) between the two terms at opposite soft signs. As $\mu_B-\mu_C=\Delta\to0$ each term grows as $1/\Delta$, while $\cM^{(N)}_-$ tends to a finite real constant, with an imaginary part linear in $\Delta$. We checked this numerically at $N=0$, 1, and 2 for $(\mu_A,\frac12(\mu_B+\mu_C))=(1,2)$, $(1.5,1.2)$, and $(0.7,2.5)$, down to $\Delta=0.01$, where at $N=1$ the sum is between $4\times10^{-3}$ and $2\times10^{-2}$ of either term (for example $\cM^{(1)}_-\to2.38\times10^{-8}$ at $(1,2)$). At equal soft masses the family is therefore a term analytic in $k_3^2$, with no $\ln k_3$, and at nearly equal masses its oscillation is slow in $\ln k_3$, with a coefficient that can exceed the pair family's.

\textit{Checks of the all-order series.}---Direct time-domain integration, with an unreleased code by the author, gives the assembled correlator at $\mu=(1,1.3,0.7)$ and $k_1=k_2=1$ on 18 values of $k_3/k$ between $0.006$ and $0.15$, with quoted errors near $10^{-12}$. After we subtract Eq.~(\ref{sm:Ssq}) through $N=8$, the remainder is a polynomial in $k_3$. A degree-seven fit leaves residuals below $1\%$ of the quoted errors, against $10\%$ when only $N=0$ is subtracted, and a free fit returns $\cM^{(1)}_+$ within $3\%$ of its closed value. For the wavefunction coefficient, where the pair family is not suppressed, we evaluate Eq.~(3) with $C_3$ from Eq.~(\ref{sm:C3iso}) at 16 values of $k_3$ between $0.03$ and $0.4$. On the 13 points with $k_3\le0.25$, the residual of a degree-eight polynomial fit falls from $1\times10^{-7}$ with nothing subtracted to $3\times10^{-14}$ with the pair series through $N=3$. Left free, $c_1$ comes out at $1.0001-0.0002\,\ii$ times its closed value and $c_2$ within $5\%$. Since $c_N$ is analytic in the orders, these checks also test the sign-flipped orders that enter the correlator. At $\mu_A=1$ the difference family at $N=0$ carries the same Boltzmann factor as the pair family and is comparable to it. The ratio $|\cM^{(0)}_-/\cM^{(0)}_+|$ is $0.78$, $0.87$, $1.22$, $1.04$, $1.25$, and $1.17$ at $(\mu_B,\mu_C)=(1.3,0.7)$, $(1.5,1)$, $(2,1)$, $(2,0.5)$, $(3,1)$, and $(4,1)$, and $4.8$ at $(3,2)$. Meanwhile $|\cM^{(0)}_-|\,e^{\pi(\mu_B+\mu_C)}$ stays between $1.7$ and $3.4\times10^{-3}$ except at $(3,2)$, where it is $1.5\times10^{-2}$. The bubble shows the same family~\cite{Aoki2026fermion}.

% ===========================================================================
% ===========================================================================
\section{Loops at strong mixing}\label{sm:strong}

We compute the pair signal of the triangle when the mixing $\rho\,\pid\sigma$ is resummed. The dressed modes are those of Refs.~\cite{Belrhali2026strong,Huenupi2026linear}, and the loop reduction is that of the main text. New here are the hard line with its time ordering, the contribution of the contact term of the interaction Hamiltonian to the one-loop bispectrum, and the continuation of the soft block. The contact term itself is in Eq.~(2.12) of Ref.~\cite{Pinol2026strong}, where it first enters the trispectrum. We set $H=1$ and $c_s=1$, and keep the cubic coupling $-\frac\alpha2\pid\sigma^2$ of Refs.~\cite{Pinol2026strong,Belrhali2026strong} with all other cubic couplings off.

\textit{Dressed modes.}---With $\Lint=a^3\rho\,\pid\sigma$ in the quadratic action the momentum of $\pi$ is $\Pi=a^3(\pid+\rho\sigma)$, so that
\begin{equation}
\pid=\Pi/a^3-\rho\,\sigma\,,\qquad M^2=m^2+\rho^2\,, \label{sm:pidot}
\end{equation}
where $M$ is the mass of $\sigma$ in the Hamiltonian. With $x=-k\eta$, $\sigma=\eta\,S$, and $\Pi=kQ/\eta$, the mode equations are
\begin{equation}
Q''+Q=\frac{\rho}{x}\,S\,,\qquad S''+\Bigl(1+\frac{M^2-2}{x^2}\Bigr)S=\frac{\rho}{x}\,Q\,, \label{sm:modeeq}
\end{equation}
with $\pi(0)=Q(0)/k$ and $\pid=x^2Q''/k$. Writing $Q=\cN\int_C\dd p\,e^{\ii xp}q(p)$ and $S=\cN\int_C\dd p\,e^{\ii xp}s(p)$, with $C$ running from $p=1$ to $1+\ii\infty$, reduces Eqs.~(\ref{sm:modeeq}) to one associated Legendre equation for $h=(1-p^2)q$, of degree $\ell=-\half+\nu_{\rm eff}$ and order $\mp\ii\rho$, with $s=(\ii/\rho)h'$. This is the Laplace-space representation of Ref.~\cite{Belrhali2026strong}. The two Bunch--Davies modes are $h_\pm=P^{\mp\ii\rho}_\ell(p)$ with $\cN_\pm=|\Gamma(1\pm\ii\rho)|\,e^{\pm\pi\rho/4}/|\Gamma(\pm\ii\rho/2)|$, fixed by the early-time behavior $e^{\ii x}x^{\mp\ii\rho/2}$. The endpoint $p=1$ is regulated by analytic continuation in the order. With these modes $2\sum_\pm|Q_\pm(0)|^2$ reproduces the exact power spectrum of Ref.~\cite{Huenupi2026linear} to 30 digits at light and heavy masses. The tree bispectra of the four couplings $\pid^3$, $\pid^2\sigma$, $\pid\sigma^2$, and $\sigma^3$ built from the same legs match the public code of Ref.~\cite{Pinol2026strong} to $2\times10^{-6}$ on five triangles at $(\rho,\mu_{\rm eff})=(1,1.5)$, $(2,2)$, and $(4,1)$, with one convention factor common to all four.

At late times the $\sigma$ component of each mode has the non-analytic terms $S_\pm\supset C^\pm_{\epsilon}\,x^{1/2+\epsilon\nu_{\rm eff}}$, $\epsilon=\pm1$, whose coefficients follow from the large-$p$ behavior $P^{\mp\ii\rho}_\ell(p)\simeq A_\pm p^\ell+B_\pm p^{-\ell-1}$ by Watson's lemma,
\begin{equation}
C^\pm_{-}=\frac{\ii\cN_\pm}{\rho}\,A_\pm\,\ell\,\Gamma(\ell)\,\ii^{\ell}\,,\qquad
C^\pm_{+}=-\frac{\ii\cN_\pm}{\rho}\,B_\pm\,(\ell+1)\,\Gamma(-\ell-1)\,\ii^{-\ell-1}\,, \label{sm:late}
\end{equation}
with $A_\pm=2^\ell\Gamma(\ell+\half)/[\sqrt\pi\,\Gamma(\ell\pm\ii\rho+1)]$ and $B_\pm=2^{-\ell-1}\Gamma(-\ell-\half)/[\sqrt\pi\,\Gamma(-\ell\pm\ii\rho)]$. A direct fit of the dressed modes at late times reproduces Eq.~(\ref{sm:late}) to $6\times10^{-6}$ at four points, two of them at $m=0$. By Eq.~(\ref{sm:pidot}) the same terms of $\pid$ are $-\rho$ times those of $\sigma$, since $\Pi/a^3$ starts two powers of $x$ later. The fitted modes confirm this to $2\times10^{-4}$ or better.

\textit{The interaction Hamiltonian.}---Because the cubic coupling contains $\pid$, the interaction Hamiltonian is not $-\Lint$. With $\mathcal{L}\supset a^3[\half\pid^2+\pid J]$ and $J=\rho\sigma-\frac\alpha2\sigma^2$, the Legendre transform gives $H\supset(\Pi-a^3J)^2/2a^3$, and therefore
\begin{equation}
H_{\rm int}=a^3\Bigl[\frac\alpha2\,\sigma^2\pid+\frac{\alpha^2}{8}\,\sigma^4\Bigr]\,, \label{sm:Hint}
\end{equation}
with $\pid$ the operator of Eq.~(\ref{sm:pidot}) and the term $\rho^2\sigma^2/2$ absorbed into $M^2$. The quartic term is the contact term of derivative couplings. Equivalently it is the $\delta$-function part of the time derivatives of the time-ordered $\pi\pi$ propagator~\cite{Abolhasani2022}. Of its 24 contractions with the two hard legs and the two soft lines, 16 come from the hard line of the triangle and 8 from the diagram in which one cubic vertex carries both hard legs. It reaches the correlator only through the $\sigma$ components of the external legs, so it vanishes as $\rho^2$ at weak mixing.

\textit{Factorization.}---In the squeezed limit $k_3\ll k_1=k_2$ the lines $B$ and $C$ and the vertex at $\bk_3$ are soft, and the hard line $A$ carries momentum $k_1$. On the rotated contour, the external leg at a vertex has a $\pid$ component $E_{\pid}(t)$ and a $\sigma$ component $E_\sigma(t)$. Summing the assignments of the fields of $\frac\alpha2\sigma^2\pid$ to the external leg, the hard line, and the soft line gives the weight
\begin{equation}
2\bigl[(E_{\pid}-\rho E_\sigma)\,A_\sigma+E_\sigma\,A_{\pid}\bigr] \label{sm:weight}
\end{equation}
at each hard vertex, where $A_X$ marks the field at that end of the hard line and the soft line enters through Eq.~(\ref{sm:late}). At the soft vertex the external leg is either $\pid$ (with $\sigma$ at both soft ends) or $\sigma$ (with $\pid$ at one soft end), with relative weight $1:2$. For the contact term all four fields are $\sigma$. We confirm these weights by brute-force enumeration of the contractions.

\textit{The soft block.}---Let $F_X(u)$ be the $X=\pid,\sigma$ component of a mode at the soft end of a line, conjugated and continued to the rotated contour, with $u$ the line energy times the vertex time. Its Mellin transform is a Gamma function times a one-dimensional integral over the Legendre weight,
\begin{equation}
\tilde F_\sigma(z)=\ii\cN\,\Gamma(z+1)\,\Omega_s(z+1)\,,\qquad \tilde F_{\pid}(z)=\cN\,\Gamma(z+2)\,\Omega_q(z+2)\,, \label{sm:Ftilde}
\end{equation}
with $\Omega_s(w)=\int_1^\infty\dd p\,p^{-w}\bar s(p)$ and $\Omega_q(w)=\int_1^\infty\dd p\,p^{2-w}\bar q(p)$. The soft block is then
\begin{equation}
\mathcal{W}(\kappa)=\iint\!\frac{\dd z_B\,\dd z_C}{(2\pi\ii)^2}\,\tilde F_B(z_B)\,\tilde F_C(z_C)\,C_2(\delta_B,\delta_C)\,\tilde E_3(-3-z_B-z_C)\,,\qquad \delta=\tfrac12\bigl(\thalf+z-\kappa\bigr)\,, \label{sm:softblock}
\end{equation}
with $C_2$ of Eq.~(5), $\tilde E_3$ the Mellin transform of the soft external leg, and $\kappa=\pm\nu_{\rm eff}$ the late-time exponent at the hard ends. No strip makes everything converge at once. The momentum integral needs $\Re(z_B+z_C)>2\Re\kappa$, and the soft vertex integral needs $\Re(z_B+z_C)<-\thalf$ ($-2$ for a $\sigma$ external leg). We therefore define $\mathcal{W}$ by continuation in $\kappa$ from $\Re\kappa\lesssim-1$, where the strips overlap. On the way to $\Re\kappa=0$ the pole of $\Gamma(\cS-\thalf)$ at $\cS=\thalf$ crosses the contour, and for a $\sigma$ external leg also the pole at $\cS=\half$. Their residues need $\tilde E_3$ outside its strip of convergence. We continue $\Omega_{s,q}$ there by integrating to $p=2$ numerically and using the convergent large-$p$ series of $P^{m}_\ell(p)$ above it, termwise in closed form. At $\Re\kappa=-\thalf$, where the momentum and time integrals converge directly, Eq.~(\ref{sm:softblock}) agrees with brute-force quadrature to $2\times10^{-4}$ or better at $(\rho,\mu_{\rm eff})=(1,1)$ and $(3,2)$, for every mode pair and both soft-vertex assignments.

\textit{The hard line.}---The hard line between the two hard vertices enters with its time ordering. On the rotated contour its time-ordered propagator is $\sum_\pm X_\pm(\ii t_<)\,\overline{Y}_\pm(\ii t_>)$, with $X$ the field at the later vertex, where $\overline{Y}$ denotes the conjugated mode continued to $\eta=\ii t$ (decaying, from the Laplace integral) and $X(\ii t)$ the mode itself (growing). We obtain $X(\ii t)$ by integrating Eqs.~(\ref{sm:modeeq}) from $\eta=-\ii t_0$, where it is the conjugate of the Laplace integral, around $\eta=-t_0$ to $\eta=\ii t_0$, and then along the imaginary axis. At the points tested the resulting Wronskians equal $-\ii$ to $3\times10^{-9}$ over $10^{-6}<t<30$, and the equal-time commutator $[\pid,\sigma]$ vanishes to $6\times10^{-9}$. The two hard time integrals are done by cumulative quadrature on all four branch pairs. The contact term needs only one hard time integral.

\textit{Checks and numerical accuracy.}---At $\rho=0.01$ the sum of the triangle and the contact term reproduces $\cM_{\rm tri}$ of the main text to $5\times10^{-4}$ at $\mu=0.7$, 1, 1.5, and 2 and at $\mu_{\rm eff}=3$, 5, 6, and 7. Refining all grids changes the strong-mixing results by at most $2\times10^{-4}$. At the largest masses the soft contour must extend to $|\Im z|\simeq2\mu_{\rm eff}+6$. Each point takes about ten seconds on one core.

\textit{Physical range.}---At fixed $\mu_{\rm eff}$ the bare mass is $m^2=\mu_{\rm eff}^2+\frac94-\rho^2$. For $m^2<0$ the Goldstone has a gradient instability (in flat space its speed squared is $m^2/(m^2+\rho^2)$), so we restrict to $\rho^2\le\mu_{\rm eff}^2+\frac94$. The endpoint $m=0$ is the case in which all of the mass of $\sigma$ comes from the mixing.

\begin{figure}[h]
\centering
\includegraphics[width=0.6\textwidth]{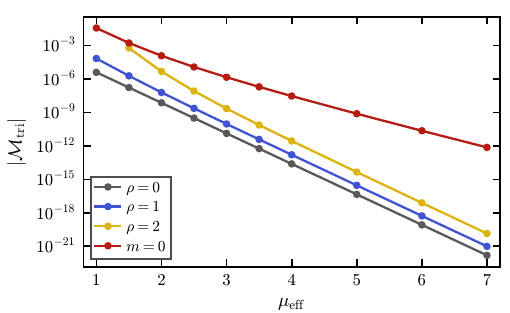}
\caption{Leading squeezed coefficient $|\cM_{\rm tri}|=|\cM^{(0)}_+|$ of the triangle with $\pid$ legs at equal masses and unit couplings, contact term included, at mixing $\rho=0$ (weak mixing), $\rho=1$ and 2, and along $m=0$, where $\rho^2=\mu_{\rm eff}^2+\frac94$. Along $m=0$ the local slope approaches $-\pi$ (it is $-3.4$ at $\mu_{\rm eff}\simeq6.5$), against $-2\pi$ at fixed $\rho$.}
\label{fig:S5}
\end{figure}

\textit{Results.}---Table~\ref{tab:strong} gives the enhancement $E=\cM_{\rm tri}(\rho)/\cM_{\rm tri}(0)$ at fixed $\mu_{\rm eff}$ and unit couplings, and the same ratio divided by $R^2$, where $R$ is the enhancement of the dressed power spectrum~\cite{Huenupi2026linear}. Its phase moves by order one, so templates built with the weak-mixing phase would be misphased at $\rho\gtrsim1$. The contact term is between 0.7 and 1.5 times the triangle at $m=0$. Along $m=0$ the local slope $\dd\ln|\cM_{\rm tri}|/\dd\mu_{\rm eff}$ is $-6.2$, $-4.6$, $-4.0$, $-3.7$, $-3.4$ near $\mu_{\rm eff}=1.25$, 2.25, 3.25, 4.5, and 6.5, approaching $-\pi$ from below. So $|\cM_{\rm tri}|$ falls as $e^{-\pi\mu_{\rm eff}}$ up to a power, far from the weak-mixing slope $-2\pi$. At fixed $\rho$, instead, the slope returns to $-2\pi$ at large $\mu_{\rm eff}$ (Fig.~\ref{fig:S5}).

\begin{table}[h]
\centering
\begin{tabular}{c|cccc|cccc}
 & \multicolumn{4}{c|}{$|E|$} & \multicolumn{4}{c}{$|E|/R^2$}\\
$\rho$ & $\mu_{\rm eff}=1$ & 2 & 3 & 5 & 1 & 2 & 3 & 5\\
\hline
0.5 & 2.44 & 2.05 & 1.96 & 1.92 & 1.82 & 1.89 & 1.90 & 1.90\\
1 & 17.4 & 8.59 & 7.02 & 6.44 & 5.04 & 5.97 & 6.15 & 6.17\\
2 & -- & 649 & 168 & 100 & -- & 79.3 & 83.4 & 83.6\\
3 & -- & -- & $1.35\times10^4$ & $2.16\times10^3$ & -- & -- & $1.06\times10^3$ & $1.34\times10^3$\\
$m=0$ & $9.3\times10^3$ & $1.7\times10^4$ & $1.1\times10^5$ & $1.7\times10^7$ & 147 & 348 & $2.5\times10^3$ & $3.8\times10^5$
\end{tabular}
\caption{Enhancement of the triangle's pair-signal coefficient (contact term included) at fixed $\mu_{\rm eff}$ and unit couplings, and the same per dressed $P_\zeta^2$. The row $m=0$ is at $\rho^2=\mu_{\rm eff}^2+\frac94$, i.e.\ $\rho=1.80$, 2.50, 3.35, and 5.22.}
\label{tab:strong}
\end{table}

\textit{The bubble.}---The same construction gives the pair signal of the bubble with a quartic vertex $\pid^2\sigma^2$ at the hard end and $\pid\sigma^2$ at the soft end, the model of Ref.~\cite{XianyuZhang2023}. Its hard vertex carries the weights $1:-2\rho:\rho^2$ for $\pid\pid$, $\pid\sigma$, and $\sigma\sigma$ external components. At $\rho=0.01$ it reproduces the weak-mixing coefficient, up to a convention factor, to $2\times10^{-4}$ at $\mu=0.7$--2. At $m=0$ it is enhanced by $3.5\times10^3$, $5.6\times10^3$, $4.1\times10^4$, and $6.8\times10^6$ at $\mu_{\rm eff}=1$, 2, 3, and 5, and its slope along $m=0$ reaches $-3.13$ at $\mu_{\rm eff}\simeq6.5$.

\textit{Which loop dominates.}---In the effective theory the coupling $\alpha$ comes from the operator $\delta g^{00}\sigma^2$~\cite{Pinol2026strong}, which also contains the quartic vertex $\pid^2\sigma^2-(\partial_i\pi)^2\sigma^2/a^2$. For the canonical field this vertex carries $\alpha/4f_\pi^2$, suppressed by $H^2/f_\pi^2\approx3\times10^{-4}$. The bubble it generates is therefore of order $\alpha^2H^2/f_\pi^2$, against $\alpha^3$ for the triangle and contact term. Up to the order-one factors of the two loop integrals, the triangle dominates unless $\alpha\lesssim H^2/f_\pi^2$.

\textit{Loop versus tree.}---The same coupling gives a tree-level bispectrum, the contact diagram with three dressed legs, whose squeezed limit oscillates as $\Re[c\,(k_3/k)^{1/2+\ii\mu_{\rm eff}}]$. We compute it from the legs used for the loop and extract $c$ by fixed-exponent fits over $10^{-3}\le k_3/k\le0.05$. The relative residuals are between $4\times10^{-11}$ and $5\times10^{-8}$, one to two orders of magnitude below those with the neighboring exponents, and narrowing the range changes $c$ by less than $10^{-5}$. At $\rho=0.05$ the tree reproduces $-\frac{\rho^2}{2}$ times the double exchange of Sec.~\ref{sm:double} summed over leg assignments to between 0.2\% and 6\% at three shapes. Table~\ref{tab:looptree} compares the oscillation amplitudes, $2|\cM_{\rm tri}|(k_3/k)^2$ for the loop and $|c|(k_3/k)^{1/2}$ for the tree, at $k_3/k=0.1$, for unit $\alpha$ and at the perturbativity bound of Ref.~\cite{Philcox2026lim}, whose Lagrangian normalizes $\alpha$ as here. The loop's normalization rests on its weak-mixing limit and a field-counting factor, so the ratios are good to about a factor of two. At fixed $\mu_{\rm eff}$ the loop's relative weight grows by a factor of 4 to 12 between $\rho=1$ and $m=0$, but it stays below $10^{-3}$ at $1\le\mu_{\rm eff}\le5$. Along $m=0$ it falls from $2\times10^{-4}$ at $\mu_{\rm eff}=1$ to $3\times10^{-8}$ at $\mu_{\rm eff}=5$. At weak mixing the tree scales as $\rho^2$ and the loop does not, so the ratio grows as $\rho\to0$. Extrapolating the trend at $\mu_{\rm eff}=1$ and $\alpha=1$, loops overtake the tree only for $\rho\lesssim10^{-2}H$, where the mixing is essentially absent. Other cubic couplings would add to the tree.

\begin{table}[h]
\centering
\begin{tabular}{ccc|cc}
$\mu_{\rm eff}$ & $\rho$ & $m^2$ & $\alpha=1$ & $\alpha=\rho^{3/4}$\\
\hline
1 & 0.25 & 3.19 & $1.1\times10^{-4}$ & $1.4\times10^{-5}$\\
1 & 0.5 & 3.0 & $4.6\times10^{-5}$ & $1.6\times10^{-5}$\\
1 & 1.0 & 2.25 & $4.6\times10^{-5}$ & $4.6\times10^{-5}$\\
1 & 1.5 & 1.0 & $7.4\times10^{-5}$ & $1.4\times10^{-4}$\\
1 & 1.8 & 0 & $1.9\times10^{-4}$ & $4.5\times10^{-4}$\\
2 & 0.25 & 6.19 & $7.5\times10^{-6}$ & $9.4\times10^{-7}$\\
2 & 1.0 & 5.25 & $2.0\times10^{-6}$ & $2.0\times10^{-6}$\\
2 & 2.0 & 2.25 & $6.3\times10^{-6}$ & $1.8\times10^{-5}$\\
2 & 2.5 & 0 & $1.0\times10^{-5}$ & $4.0\times10^{-5}$\\
3 & 1.0 & 10.25 & $1.0\times10^{-7}$ & $1.0\times10^{-7}$\\
3 & 2.0 & 7.25 & $2.0\times10^{-7}$ & $5.7\times10^{-7}$\\
3 & 3.0 & 2.25 & $9.2\times10^{-7}$ & $4.8\times10^{-6}$\\
3 & 3.35 & 0 & $1.2\times10^{-6}$ & $7.3\times10^{-6}$\\
5 & 5.22 & 0 & $2.9\times10^{-8}$ & $3.5\times10^{-7}$
\end{tabular}
\caption{Ratio of the oscillation amplitudes of the loop (triangle and contact term, at frequency $2\mu_{\rm eff}$) and of the tree (at $\mu_{\rm eff}$) in the bispectrum of the coupling $-\frac\alpha2\pid\sigma^2$, at $k_3/k=0.1$, for unit $\alpha$ and at the perturbativity bound $\alpha=(\rho/H)^{3/4}$ of Ref.~\cite{Philcox2026lim}. The ratio scales as $\alpha^2$ and as $(k_3/k)^{3/2}$.}
\label{tab:looptree}
\end{table}

\end{document}